\documentstyle[12pt]{article}
\renewcommand{\today}{4 April 1996}
\makeatletter
\@ifundefined{selectfont}{
  \def\mathrm#1{{\rm #1}}
}{
  \let\oldmathrm=\mathrm
  \def\mathrm#1{{\oldmathrm{#1}}}
}
\@ifundefined{reset@font}{\let\reset@font\empty}{}

\def\Bbb#1{{\bf\relax#1}}
\def\section{\@startsection {section}{1}{\z@}{-3.5ex plus-1ex minus
    -.2ex}{2.3ex plus.2ex}{\reset@font\normalsize\bf\boldmath}}
\def\subsection{\@startsection{subsection}{2}{\z@}{-3.25ex plus-1ex
    minus-.2ex}{1.5ex plus.2ex}{\reset@font\normalsize\it}}
\def\subsubsection{\@startsection{subsubsection}{3}{\z@}{-3.25ex plus
 -1ex minus-.2ex}{1.5ex plus.2ex}{\reset@font\normalsize\bf}}
\advance\textheight by \topskip
\makeatother
\input amssym.def       
\input amssym.tex       
\def\Cop{{\Bbb C}}
\def\Zop{{\Bbb Z}}

\def\Nop{{\Bbb N}}
\def\bbbone {{\mathchoice {\rm 1\mskip-4mu l} {\rm 1\mskip-4mu l}
{\rm 1\mskip-4.5mu l} {\rm 1\mskip-5mu l}}}
\def\tr{\mathop{\rm Tr}\nolimits}

\def\Vir{{\mathop{\sf Vir}}}
\def\multirow#1#2{\count0=#1 \advance\count0 by -1
  \dimen0=\ht\strutbox\advance\dimen0\dp\strutbox
  \dimen0=\count0\dimen0
  \setbox2=\hbox{\raise0.5\dimen0\hbox{$#2$}}%
  \ht2=\ht\strutbox \dp2=\dp\strutbox \box2}
\def\multiparen#1#2#3{\setbox2=\null \ht2=#1\ht\strutbox \dp2=#1\dp\strutbox
  \setbox2=\hbox{$\left#2\vcenter{\box2}\right#3$}
  \dimen0=\dp2 \advance\dimen0-\dp\strutbox
  \setbox2=\hbox{\raise\dimen0\box2}
  \ht2=\ht\strutbox \box2}
\makeatletter

\@addtoreset{equation}{section}
\makeatother
\begin{document}
\thispagestyle{empty}
\begin{flushright}
  hep-th/9604026\\
DAMTP 96-36
\end{flushright}
\vskip 2em
\begin{center}\LARGE
  Indecomposable Fusion Products
\end{center}\vskip 1.5em
\begin{center}\large
  Matthias R. Gaberdiel\footnote{Email: {\tt M.R.Gaberdiel@damtp.cam.ac.uk}}
  and
  Horst G. Kausch\footnote{Email: {\tt H.G.Kausch@damtp.cam.ac.uk}}
\end{center}
\begin{center}\it
Department of Applied Mathematics and Theoretical
Physics, \\
University of Cambridge, Silver Street, \\
Cambridge CB3 9EW, U.K.
\end{center}
\vskip 1em
\begin{center}
  \today
\end{center}
\vskip 1em
\begin{abstract}
  We analyse the fusion products of certain representations of the
  Virasoro algebra for $c=-2$ and $c=-7$ which are not completely
  reducible.  We introduce a new algorithm which allows us to study
  the fusion product level by level, and we use this algorithm to
  analyse the indecomposable components of these fusion products. They
  form novel representations of the Virasoro algebra which we describe
  in detail.

  We also show that a suitably extended set of representations closes
  under fusion, and indicate how our results generalise to all $(1,q)$
  models.

\end{abstract}

\section{Introduction}
\label{sec:intro}

Recently it has become apparent that there exist conformal field
theories whose correlation functions exhibit logarithmic behaviour.
The presence of the logarithmic terms has been interpreted to signal
the appearance of a new type of conformal operator, now usually called
{\it logarithmic operator}.  Models with such logarithmic operators
include the WZNW model on the supergroup $GL(1,1)$ \cite{RSal92}, the
$c=-2$ model \cite{Gur93}, gravitationally dressed conformal field
theories \cite{BKog95} and some critical disordered models
\cite{CKT95}.
They are believed to be important for the description of certain
statistical models, in particular in the theory of (multi)critical
polymers \cite{Sal92,Flohr95,Kau95} and percolation \cite{Watts96}. 
There have also been suggestions that some of these logarithmic
operators might correspond to normalisable zero modes for string
backgrounds \cite{KMa95}.

It was already realised in \cite{Gur93} that the logarithmic
operators are not eigenstates of the scale generator $L_0$, but that,
under the action of $L_0$, they form a Jordan cell with another field
of the same conformal dimension. In particular, this means that the
scale generator $L_0$ is not diagonalisable, and that the space of
states of the theory is not a direct sum of irreducible Virasoro
highest weight representations. This has raised the question of how to
understand these novel representations of the Virasoro algebra
\cite{Flohr95,Kau95}, and it is one of the aims of this paper to
give a detailed description of a certain class of such indecomposable
representations.
\smallskip

The way in which the logarithmic operators appear in the theory is as
follows. It is well known that the two-- and three--point functions of
a (chiral) conformal field theory are determined up to constants by
the conformal symmetry, and that every four--point function is fixed
up to an {\it a priori} undetermined function $f(x)$ of one (complex)
variable, the cross-ratio of the four coordinates. There is a subclass
of theories (the so-called {\it degenerate} theories), for which
certain descendents of the fields decouple, and this gives rise to
differential equations for the undetermined function $f(x)$. The
differential equations have typically (regular) singular points at
$x=0,1$ and $\infty$, and we can thus solve them by a power series
expansion about any one of the singular points. In general, this will
give all (different) solutions to the differential equation, unless
two of the leading powers differ by an integer. Then there will
sometimes be one (or more) solutions which involve a logarithmic
term. Typically we cannot ignore these logarithmic solutions, as this
would destroy the crossing symmetry of the conformal field theory.  

The fields of a conformal field theory can be identified with states
in a representation of the symmetry algebra, and the product of
fields, the so-called {\it fusion}, can be understood as some kind of
tensor product of these representations (which is again a
representation of the symmetry algebra) \cite{Gab93,Nahm94}. The
different solutions to the differential equation for $f(x)$ (near the
singular point $x=0$, say) correspond then to the different
subrepresentations which are contained in the fusion product (of the
two fields whose distance vanishes at $x=0$). In the usual case, the
action of $L_0$ on this tensor product is diagonalisable, and the
product can be decomposed into a direct sum of irreducible
representations.  In the logarithmic case, however, the action of
$L_0$ is no longer diagonalisable, and the product contains reducible
but indecomposable representations \cite{Nahm94}.
\medskip

In this paper we want to study the indecomposable representations
which are contained in the fusion products of certain highest weight
representations of the Virasoro algebra for $c=-2$ and $c=-7$. We
introduce a new algorithm which allows us to study the fusion product
up to any (finite) level while only considering finite dimensional
vector spaces. We then use this algorithm to study the fusion product
level by level, and we can thus unveil the structure of the
indecomposable representations.

It turns out that only certain types of indecomposable representations
appear in the various fusion products, and we describe them in detail.
These representations are characterised by one parameter which we
determine explicitly for a number of cases.  Similar representations
have also been investigated independently by Rohsiepe \cite{Roh96}.  

We analyse the fusion of these indecomposable representations,
and find that a certain set of representations, containing all highest
weight representations and some indecomposable representations, is
again closed under fusion. Finally, we indicate how some of our
results generalise naturally to all $(1,q)$ models. 
\pagebreak

The paper is organised as follows. In Section \ref{sec:fusion}, we
introduce our algorithm, and show that it always terminates for the
case of the Virasoro algebra. We then recall in Section \ref{sec:vir},
what is known about the fusion of highest weight representations of
the Virasoro algebra, and explain why it is natural to expect that
indecomposable representations appear in certain fusion products of
the $(1,q)$ models. We describe the structure of the indecomposable
representations in Section \ref{sec:reps}, and give a conjecture for
the general fusion rules. In Section \ref{sec:results} we present
the explicit results of our calculations which provide supportive
evidence for these conjectures. In Section \ref{sec:disc}, we make
some prospective remarks, and in the appendix we give explicit details
of one specific example to illustrate our method.

\section{Fusion and tensor products: The algorithm}
\label{sec:fusion}

Let us start by introducing some notation. We denote by
${\cal A}$ the chiral algebra, {\it i.e.} the algebra
generated by the modes of the holomorphic fields, and fix our
convention as in  \cite{Goddard89}, so that the modes of a field $S$
of conformal weight $h$ are given as
\begin{equation}
\label{mode}
S(w)=\sum_{l\in\Zop + h} w^{l-h} \; S_{-l} \,.
\end{equation}
We also assume, as is usual in conformal field theory, that one of the
fields is the stress-energy tensor $L(z)$ of weight $2$, whose modes
$L_{n}$ satisfy the Virasoro algebra.

Given two representations of ${\cal A}$, ${\cal H}_1$ and ${\cal H}_2$,
and two points $z_1, z_2 \in \Cop$ in the complex plane, 
the fusion tensor product can be defined by
the following construction \cite{Gab93}. First we consider the product
space $\left( {\cal H}_1 \otimes {\cal H}_2 \right)$ on which two
different actions of the chiral algebra are given by the two
comultiplication formulae \cite{Gab94}
\begin{eqnarray}
  {\displaystyle \Delta_{\zeta,z}(S_{n}) =
    \widetilde{\Delta}_{\zeta,z}(S_{n})} = &  
  {\displaystyle \sum_{m=1-h}^{n} \left( \begin{array}{c} n+h-1 \\ m+h-1
  \end{array} \right)
  \zeta^{n-m} \left(S_{m} \otimes \bbbone\right)  }
\hspace*{4cm} \nonumber \\
\label{chir1}
&  \hspace*{3cm} {\displaystyle +\, \varepsilon_{1}
  \sum_{l=1-h}^{n} \left( \begin{array}{c} n+h-1 \\ l+h-1
\end{array} \right)
z^{n-l} \left(\bbbone \otimes S_{l} \right)\,,}  \\
{\displaystyle \Delta_{\zeta,z}(S_{-n})} = &
{\displaystyle \sum_{m=1-h}^{\infty} \left( \begin{array}{c} n+m-1 \\ n-h
\end{array} \right) (-1)^{m+h-1}
\zeta^{-(n+m)} \left(S_{m} \otimes \bbbone \right) }
\hspace*{2cm} \nonumber \\
\label{chir2}
&  \hspace*{3cm} {\displaystyle +\, \varepsilon_{1}
\sum_{l=n}^{\infty} \left( \begin{array}{c} l-h \\ n-h
\end{array} \right)
(-z)^{l-n} \left(\bbbone \otimes S_{-l}  \right)\,,}
\end{eqnarray}
and
\begin{eqnarray}
\label{chir2'}
{\displaystyle \widetilde{\Delta}_{\zeta,z}(S_{-n})} = &
{\displaystyle 
\sum_{m=n}^{\infty} \left( \begin{array}{c} m-h \\ n-h
\end{array} \right) 
(-\zeta)^{m-n} \left(S_{-m} \otimes \bbbone \right) }
\hspace*{2.5cm} \nonumber \\
& \hspace*{1.5cm} {\displaystyle + \varepsilon_{1}
\sum_{l=1-h}^{\infty} \left( \begin{array}{c} n+l-1 \\ n-h
\end{array} \right) (-1)^{l+h-1}
z^{-(n+l)} \left(\bbbone \otimes S_{l}  \right)\,,}
\end{eqnarray}
where in (\ref{chir1}) we have $n\geq 1-h$, in
(\ref{chir2},\,\ref{chir2'}) $n\geq h$, and $\varepsilon_{1}$ is $\mp 1$
according to whether the left-hand vector in the tensor product
and the field $S$ are both fermionic or not.\footnote{The second
formula differs from the one given in \cite{Gab94} by a different
$\varepsilon$ factor. There the two comultiplication formulae were
evaluated on different branches; this is corrected here.} 

The fusion tensor product is then defined as the quotient of the
product space by all relations which come from the equality of
$\Delta_{z_1,z_2}$ and $\widetilde{\Delta}_{z_1,z_2}$
\begin{equation}
\left( {\cal H}_1 \otimes {\cal H}_2 \right)_{\mathrm f} :=
\left( {\cal H}_1 \otimes {\cal H}_2 \right) / 
(\Delta_{z_1,z_2} - \widetilde{\Delta}_{z_1,z_2}) \,.
\end{equation}
It has been shown for a number of examples that this definition
reproduces the known restrictions for the fusion rules
\cite{Gab93,Gab94}.

In these calculations, the ring-like nature of the fusion product was
not really used, and only certain necessary conditions for the fusion
rules were derived. In this paper we want to introduce (and use) a
more refined method for the study of the fusion product. We shall
define a family of typically finite-dimensional quotient spaces which
provide a filtration for the whole space. The action of the chiral
algebra can be studied on all of these spaces, and it will turn out
that essentially the whole structure of the fusion product can be
understood by considering a finite number of them. This analysis is a
natural generalisation of some ideas of Werner Nahm \cite{Nahm94}.

To define this filtration let us introduce some more notation. The
chiral algebra ${\cal A}$ contains two rather natural subalgebras.
Firstly, there is the algebra which is generated by the negative modes
which annihilate the vacuum
\begin{equation}
{\cal A}^0_{-} : = \langle S_{-n} | 0 < n < h(S) \rangle \,,
\end{equation}
and secondly, the algebra generated by the negative modes which do not
annihilate the vacuum
\begin{equation}
{\cal A}_{--} : = \langle S_{-n} | n \geq h(S) \rangle \,.
\end{equation}
We also denote by ${\cal A}_{-}$ and ${\cal A}_{+}$ the algebras
generated by all negative and positive modes, respectively.

Let ${\cal H}$ be a representation of the chiral algebra ${\cal A}$. 
We define the {\it special subspace} of ${\cal H}$ as the quotient
space\footnote{We should note that ``the special subspace'' 
is not defined as a subspace.} \cite{Nahm94}
\begin{equation}
{\cal H}^{\mathrm{s}} : = {\cal H} / {\cal A}_{--} {\cal H} \,.
\end{equation}
If ${\cal H}$ is a highest weight representation,
{\it i.e.} if ${\cal H}$ is generated by the action of ${\cal
A}_{-}$ from a highest weight vector $\psi$ which satisfies ${\cal
A}_{+}\psi=0$, then it is always possible to realise the special
subspace as a quotient space of ${\cal A}^0_{-} \psi$. If not
explicitly mentioned otherwise, we shall use this convention in the
following. For the case of the Virasoro algebra (to which we shall
restrict ourselves for a large part of the paper), there exists only
one realisation of the special subspace as a subspace of 
${\cal A}^0_{-} \psi$. 

In the following we shall only consider {\it quasirational}
representations, {\it i.e.} representations for which the dimension of
the special subspace is finite. As shown in \cite{Nahm94},
the fusion product of a quasirational representation with a highest
weight representation contains at most finitely many (irreducible) 
subrepresentations. 

The special subspace is a very important concept for the analysis of
the fusion product, but it has the rather important drawback that it
does not carry a representation of ${\cal A}_{-}^{0}$ or even the zero
modes. (For example, in the case of the $W_3$ algebra \cite{FZ87}, the
action of $W_0$ is not defined on ${\cal H}^{\mathrm{s}}$, as
$[W_0,L_{-2}] = 4 W_{-2}$.) It is therefore useful to introduce a
different set of quotient spaces. We define the subalgebra which is
generated by all products of modes whose $L_0$ grading is greater or
equal to $n$,
\begin{equation}
{\cal A}_{n} : = \langle \prod_{j=1}^{m} S^{k_j}_{-l_j} |
\sum_{j=1}^{m} l_j \geq n \rangle\,,
\end{equation}
and define then a filtration of ${\cal H}$ as the family of quotient
spaces
\begin{equation}
{\cal H}^{n} : = {\cal H} / {\cal A}_{n+1} {\cal H} \,.
\end{equation}
In particular, for a highest weight representation, ${\cal H}^{0}$ can
be naturally identified with the space of highest weight states, and
${\cal H}^n$ with the space of all descendents up to level $n$.
However, the advantage of this description is that it can equally be
applied to representations which are not necessarily highest weight
representations.

The important insight of Nahm in \cite{Nahm94} was that
\begin{equation}
\left( {\cal H}_1 \otimes {\cal H}_2 \right)_{\mathrm f}^{0} \subset
{\cal H}_1^{\mathrm{s}} \otimes {\cal H}_2^{0} \hspace*{1cm}
\mbox{and} \hspace*{1cm}
\left( {\cal H}_1 \otimes {\cal H}_2 \right)_{\mathrm f}^{0} \subset
{\cal H}_1^0 \otimes {\cal H}_2^{s} \,.
\end{equation}
This was shown by giving an algorithm for reducing the left-hand-side,
using the two comultiplications. We want to conjecture here (and give
a proof for the case of the Virasoro algebra), that actually much more
is true, namely 
\begin{equation}
\label{conjecture}
\left( {\cal H}_1 \otimes {\cal H}_2 \right)_{\mathrm f}^{n} \subset
{\cal H}_1^{\mathrm{s}} \otimes {\cal H}_2^{n} \hspace*{1cm}
\mbox{and} \hspace*{1cm}
\left( {\cal H}_1 \otimes {\cal H}_2 \right)_{\mathrm f}^{n} \subset
{\cal H}_1^n \otimes {\cal H}_2^{s} \,.
\end{equation}
The strategy to show this is again to give an algorithm for reducing
the left-hand-side. We have been unable so far to show that the
algorithm terminates in general. However, for the Virasoro algebra which
we shall study for most of the rest of the paper it is easy to see
that it does.  

The importance of this result is that it enables one to analyse the
fusion product level by level, in a way which respects the action of
the modes. For example, the comultiplication formula gives an action
of $S_{m}$,
\begin{equation}
S_{m}: \left( {\cal H}_1 \otimes {\cal H}_2 \right)_{\mathrm f}^{n}
\rightarrow 
\left( {\cal H}_1 \otimes {\cal H}_2 \right)_{\mathrm f}^{n-m} 
\end{equation}
for $m<n$. In particular, the zero modes map
each space into itself, and we can thus analyse the eigenvalues and
eigenvectors up to any level.  

For the application we have primarily in mind --- the analysis of not
completely reducible representations of the Virasoro algebra --- it is
essential to be able to analyse all of these spaces. In
particular as we shall see, there exist not completely reducible
representations which seem to be completely reducible if only the
lowest energy space of the fusion product is analysed in the way
proposed by Nahm \cite{Nahm94}. The method used here not only shows
that they are in fact not completely reducible, but it also allows one
to calculate  certain {\it a priori} free parameters which
characterise different representations of the same general type. 

More fundamentally, (\ref{conjecture}) gives an upper bound on the
number of states of a certain level in the fusion product. In
particular, it reaffirms the interpretation of Nahm about the
dimension of the special subspace being closely related to the quantum
dimension of the corresponding representation. It should also have
implications for questions of convergence of characters. 
\medskip

As mentioned before, the strategy for proving (\ref{conjecture}) is to
give an algorithm for reducing the left-hand-side. We want to explain
this algorithm now in more detail. To fix notation let us assume (for
simplicity) that the two parameters are $(z_1,z_2)=(1,0)$. We shall
only consider the proof for the first case, the other case being
completely analogous because of symmetry. The algorithm consists
essentially of two steps which are applied alternately.
\bigskip

{\leftskip=1truecm

{\bf (A1)}: Rewrite a given vector 
\begin{equation}
\psi_1 \otimes \psi_2 \in {\cal H}_1 \otimes {\cal H}_2 
\end{equation}
as
\begin{equation}
\label{A1}
\psi_1 \otimes \psi_2 = \sum_i \psi^i_1 \otimes \psi^i_2 + 
\Delta_{1,0}({\cal A}_{n+1}) 
\left({\cal H}_1 \otimes {\cal H}_2 \right) \,,
\end{equation}
where $\psi_1^i \in {\cal H}_1^{\mathrm{s}}$. 
\bigskip

{\bf (A2)}: Rewrite 
$\psi_1 \otimes \psi_2 \in {\cal H}_1^{\mathrm{s}} \otimes {\cal H}_2$ as
\begin{equation}
\label{A2}
\psi_1 \otimes \psi_2 = \sum_i \psi^i_1 \otimes \psi^i_2 + 
\Delta_{1,0}({\cal A}_{n+1}) 
\left({\cal H}_1 \otimes {\cal H}_2 \right) \,,
\end{equation}
where $\psi_1^i \in {\cal A}_{-}^0 {\cal H}_1^{\mathrm{s}}$ and 
$\psi_2^i \in {\cal H}_2^n$

}
\bigskip

The first step {\bf (A1)} requires some comment. We first rewrite
$\psi_1\in{\cal H}_1$ as
\begin{equation}
  \psi_1 = \sum_{j} \psi_j^{\mathrm{s}} + \sum_{k} {\cal A}_{--}
  \chi_k^{\mathrm{s}} \,,
\end{equation}
where $\psi_j^{\mathrm{s}}$ and $\chi_k^{\mathrm{s}}$ are in ${\cal
  H}_1^{\mathrm{s}}$. To rewrite the action of ${\cal A}_{--}$ on the
left-hand vector in the fusion tensor product, we use the following
crucial property which follows from \cite{Gab93,Gab94},
\begin{eqnarray}
\widetilde{\Delta}_{0,-1} (S_{-m}) & = & 
\left( e^{L_{-1}} \otimes e^{L_{-1}} \right) 
\Delta_{1,0} (S_{-m}) 
\left( e^{- L_{-1}} \otimes e^{ -L_{-1}} \right) \nonumber \\
& = & \Delta_{1,0} (e^{ L_{-1}} S_{-m} e^{- L_{-1}} ) \nonumber \\
& = & \sum_{l=m}^{\infty} 
\left( \begin{array}{c} l-h \\ m-h \end{array} \right)
\Delta_{1,0} (S_{-l}) \nonumber \\
& = & \sum_{l=m}^{n} 
\left( \begin{array}{c} l-h \\ m-h \end{array} \right)
\Delta_{1,0} (S_{-l}) + \Delta_{1,0}({\cal A}_{n+1}) \,,
\end{eqnarray}
where $h$ is the conformal weight of $S$, and we have assumed that
$m\leq n$, as otherwise the whole expression is in the subspace by
which we quotient. We note that for $m\leq l \leq n$,
$\Delta_{1,0}(S_{-l})$ is of the form ${\cal A}_{-}^{0}\otimes\bbbone 
+ \bbbone\otimes {\cal A}_{--}$.  Furthermore, we have
\begin{equation}
\widetilde{\Delta}_{0,-1} (S_{-m}) = 
\left( S_{-m} \otimes \bbbone \right) + \varepsilon_{1}
\sum_{l=1-h}^{\infty} 
\left( \begin{array}{c} m+l-1 \\ m-h \end{array} \right)
(-1)^{l+h-1} \left(\bbbone \otimes S_{l} \right) \,,
\end{equation}
and we can thus use these equations to rewrite
$(S_{-m}\otimes\bbbone)$ in terms of ${\cal A}_{-}^{0}$ acting on the
left hand vector (and some modes acting on ${\cal H}_2$). We then
rewrite these terms again as vectors in ${\cal H}_1^{\mathrm{s}}$ plus
terms of the form ${\cal A}_{--} \phi_l^{\mathrm{s}}$, where
$\phi_l^{\mathrm{s}}\in{\cal H}_1^{\mathrm{s}}$ and repeat the
procedure. Using the same argument as in \cite{Nahm94}, it is easy to
see that this algorithm always stops, as the conformal weight of the
relevant vectors decreases in each step.

To understand how the second step {\bf (A2)} can be implemented, we
note that for a monomial of negative modes $S_{-I}$ of level $|I|$,
the comultiplication formulae (\ref{chir1}) and (\ref{chir2}) imply that
\begin{equation}
\Delta_{1,0} (S_{-I}) = \left( \bbbone \otimes S_{-I}\right) +
\sum_{k} {\cal A}_{-}^0 \otimes P^{(k)}_{- I_k} \,,
\end{equation}
where $P^{(k)}_{- I_k}$ are monomials with $|I_k| < |I|$. Using
this relation repeatedly, it is clear that {\bf (A2)} can be
implemented. 
\smallskip

We want to explain now how {\bf (A1)} and {\bf (A2)} can be used to
reduce the left hand side of (\ref{conjecture}). We first apply {\bf
(A1)} to replace all terms in the left hand vector which are not
in the special subspace. We then use {\bf (A2)} to remove all states
of level higher than $n$ from the right hand vector, in the course of
which we potentially create left hand vectors which are outside 
the special subspace. We thus repeat step {\bf (A1)}, thereby
obtaining terms where arbitrary modes up to (negative) level $n$ act
on the right hand vectors. Thus, potentially we have to repeat step
{\bf (A2)}, and so on.

In order to show that the algorithm stops one could try to find a
real-valued function of the two vectors which decreases in each
cycle. Unfortunately, it is not clear how to define such a function in
general as the specific form of the null-vector (which is used to
rewrite a vector in ${\cal A}_{-}^0 \psi^0$, where $\psi^0$ is the
highest weight vector, as an element in the special subspace and
vectors of the form ${\cal A}_{--} \psi$) is crucial in step {\bf
(A1)}. We do not know general properties of these null-vectors and
thus have not been able to show the result in general.

On the other hand for the case of the Virasoro algebra, 
we can define the decreasing function to be the sum of the number of
modes of $L$ on both vectors. It is clear that this function does not
increase in step {\bf (A2)}, and in step {\bf (A1)} it has to decrease
when we replace an element in $L_{-1}^n \psi^0$ by vectors involving
$L_{-2}$ or higher modes. A similar construction also works for certain
representations of the $W_3$ algebra.

\section{Virasoro fusion revisited}
\label{sec:vir}

Let us now specifically consider the Virasoro algebra $\Vir$ which 
is generated by the modes $L_n, n\in\Bbb{Z}$ of the stress tensor with
commutation relations
\begin{equation} 
  [ L_m, L_n ] = (m-n) L_{m+n} + {c\over12} m (m^2-1) \delta_{m+n}\,.
\end{equation}
The representations of this algebra which are most relevant in
conformal field theory are {\it highest weight representations} of
definite highest weight $h$; these are generated by the creation modes
$L_n, n<0$ from a highest weight vector $v_{h}$, satisfying
\begin{equation}
\label{eq:hwv}
\begin{array}{lcl}
  L_0 v_{h} &=& h v_{h}\,, \\
  L_{n} v_{h} &=& 0, \qquad\hbox{for $n>0$}\,. 
\end{array}
\end{equation}
Here we assume in particular that the highest weight vector is a
cyclic vector with respect to the action of the Virasoro algebra, {\it
i.e.}\ that all vectors in the representation space can be obtained by
the action of the Virasoro algebra from the highest weight vector. We
shall encounter later on representations which contain a vector of
lowest $L_0$-eigenvalue, satisfying (\ref{eq:hwv}), but for which this
vector is not cyclic. These representations shall be called {\it
generalised highest weight representations}.
\smallskip

We parametrise the central charge of the Virasoro algebra as
\begin{equation}
  c = 13 - 6(t + t^{-1})\,, 
\end{equation}
and the conformal weights of highest weight vectors 
\begin{equation}
\label{eq:hparam}
  h = {\alpha^2\over4t} - {(t-1)^2\over4t}\,.
\end{equation}

The highest weight representation ${\cal M}_h$ which is freely
generated from $v_h$ by the action of
\begin{equation}
\Vir_- := \langle L_n | n<0 \rangle
\end{equation}
is called a Verma module. It may contain a non-zero 
highest weight vector $w$ of weight $h+n$, and such a vector is then 
called a {\em singular vector\/} of degree $n$. It
generates a subrepresentation of ${\cal M}_h$, and conversely all
subrepresentations of ${\cal M}_h$ are generated from highest weight
vectors.

The Verma module ${\cal M}_{h}$ has a singular vector of degree $N=mn$
if $h=h(\alpha_{m,n})$, where $\alpha_{m,n} = mt-n$ and
$m$ and $n$ are positive integers. In this case we will denote the
corresponding Verma module by ${\cal M}_{m,n}$. 

If $t$ is an integer all singular vectors of ${\cal M}_{h}$ are of
this type. Furthermore, as $\alpha_{m,n}= \alpha_{m+1, n+t}$
and $h(\alpha_{m,n})=h(\alpha_{-m,-n})$, we can always reduce the 
label $n$ to lie in the {\em fundamental domain\/}, $0<n\leq t$. Then 
\begin{displaymath}
  h_{m,n} + mn = h_{m,-n} = h_{m+1, t-n}
\end{displaymath}
and thus the subrepresentation generated by the singular vector
possesses again a singular vector. For integer $t$ we hence obtain 
the following chains of Verma module embeddings:
  
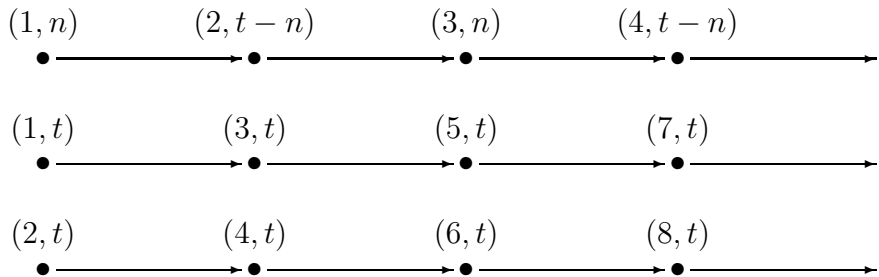
\begin{center}
    \begin{picture}(320,120)(0,-10)
      \multiput(0,80)(80,0){4}{\vbox to 0pt
        {\vss\hbox to 0pt{\hss$\bullet$\hss}\vss}}
      \multiput(5,80)(80,0){4}{\vector(1,0){70}}
      
      \multiput(0,40)(80,0){4}{\vbox to 0pt
        {\vss\hbox to 0pt{\hss$\bullet$\hss}\vss}}
      \multiput(5,40)(80,0){4}{\vector(1,0){70}}
      
      \multiput(0,0)(80,0){4}{\vbox to 0pt
        {\vss\hbox to 0pt{\hss$\bullet$\hss}\vss}}
      \multiput(5,0)(80,0){4}{\vector(1,0){70}}
      
      \put(0,90){\hbox to 0pt{\hss$(1,n)$\hss}}
      \put(80,90){\hbox to 0pt{\hss$(2,t-n)$\hss}}
      \put(160,90){\hbox to 0pt{\hss$(3,n)$\hss}}
      \put(240,90){\hbox to 0pt{\hss$(4,t-n)$\hss}}
      
      \put(0,50){\hbox to 0pt{\hss$(1,t)$\hss}}
      \put(80,50){\hbox to 0pt{\hss$(3,t)$\hss}}
      \put(160,50){\hbox to 0pt{\hss$(5,t)$\hss}}
      \put(240,50){\hbox to 0pt{\hss$(7,t)$\hss}}
      
      \put(0,10){\hbox to 0pt{\hss$(2,t)$\hss}}
      \put(80,10){\hbox to 0pt{\hss$(4,t)$\hss}}
      \put(160,10){\hbox to 0pt{\hss$(6,t)$\hss}}
      \put(240,10){\hbox to 0pt{\hss$(8,t)$\hss}}
    \end{picture}
  \end{center}
Here, a vertex with a label $(m,n)$ denotes a highest weight vector of
weight $h_{m,n}$, and an arrow $1\longrightarrow2$ means that 
the vertex $2$ is the singular vector of lowest degree contained in
the representation generated by $1$.  

In the following we shall also need the Verma module embedding
diagrams for irrational $t$. In this case there are two types of
diagrams: those corresponding to Verma modules where the highest
weight vector is described by $m,n\in\Nop$ and where there exists
precisely one singular vector of weight $h_{m,-n}$ (at level $mn$),
and those for which this is not the case, and which do not have any
singular vectors at all.
\smallskip

Any highest weight representation can be obtained as the quotient of a
Verma module by a subrepresentation. The singular vectors which
generate the corresponding subrepresentation are then called {\em null
vectors\/}. For integer $m,n\in\Nop$, we denote by ${\cal V}_{m,n}$
the highest weight representation of weight $h_{m,n}$ which is
obtained from the Verma module ${\cal M}_{m,n}$ by dividing out the
subrepresentation generated from the singular vector at level $mn$
(which is then null). For irrational $t$, all such representations are
irreducible, and for integer $t$, precisely those are irreducible for
which $n$ is in the fundamental domain $0<n\leq t$. (For $n>t$, the
Verma module ${\cal M}_{m,n}$ is the same as ${\cal M}_{m',n'}$, where
$n'$ is in the fundamental domain; ${\cal M}_{m,n}$ therefore
possesses a singular vector of level $m' n' < m n$, and ${\cal
V}_{m,n}$ is not irreducible.)

All these representations are quasirational, and hence their fusion
product decomposes into finitely many indecomposable quasirational
subrepresentations. Using the techniques of the previous section we
can analyse their fusion, and we shall do so later on in some detail.
First, however, we want to see, what can be learned about these fusion
products using the general results of Feigin and Fuchs
\cite{FFu90,FFu88}. From their point of view, the fusion of
$\phi_1\in{\cal V}_{1}$ and $\phi_2\in{\cal V}_{2}$ is the set of all
representations ${\cal P}_{3}$, for which there exists $\phi_3\in{\cal
P}_{3}$, such that the three-point function
$\langle\phi_3(z_3)\phi_2(z_2)\phi_1(z_1)\rangle$ does not vanish.
(Here $\phi(z)$ denotes the field corresponding to the state
$\phi\in{\cal P}$.) Using the M\"obius invariance of the three-point
functions, it is sufficient to consider the case, where the three
points $z_1, z_2, z_3$ are $0,1,\infty$. We can furthermore identify
fields and states at $z=0$.

Suppose now that $\phi_i\equiv\phi_{m_i,n_i}\in{\cal V}_{m_i,n_i}$ are
primary fields, and that $m_1, m_2, n_1, n_2\in\Nop$. We want to use
the fact that the three-point function has to vanish if we consider
the null descendent ${\cal N}_1$ of $\phi_1$ at level $m_1 n_1$. This
gives the equation
\begin{equation}
\label{eq:threepoint}
  0 = \langle\phi_3(\infty)\phi_2(1){\cal N}_1(0)\rangle = 
  p_{m_1,n_1}(h_2, h_3) 
  \langle\phi_3(\infty)\phi_2(1)\phi_1(0)\rangle\,,
\end{equation}
where $p_{m_1,n_1}$ factorises as \cite{FFu88} 
\begin{equation}
  p_{m_1,n_1}(h_2, h_3) \propto \prod_{r,s} 
  (\alpha_2 + \alpha_3 - \alpha_{r,s}) 
  (\alpha_2 - \alpha_3 - \alpha_{r,s})\,.
\end{equation}
Here we use the parametrisation (\ref{eq:hparam}), and the product is over 
the range 
\begin{equation}\label{eq:range}
\begin{array}{r@{}l}
  r \in{}& \{ -(m_1-1), -(m_1-3), \ldots, (m_1-1) \}, \\
  s \in{}& \{ -(n_1-1), -(n_1-3), \ldots, (n_1-1) \}\,.
\end{array}
\end{equation}
We can also use the null-vector relations for $\phi_2$, to obtain
a formula similar to (\ref{eq:threepoint}). In order for the fusion of
$\phi_1$ and $\phi_2$ to contain the highest weight representation
generated from $\phi_3$, the three-point function of some states
in the three representations has to be non-trivial. It is easy to see
that this can only be the case if the three-point function of the
corresponding highest weight vectors does not vanish, and this gives
rise, by (\ref{eq:threepoint}), to the (unintersected) fusion rules
\begin{equation}
\label{eq:fusion-rule}
  \phi_{m_1,n_1} \times \phi_{m_2,n_2} = 
        \sum_{m_3=|m_1-m_2|+1}^{m_1+m_2-1} 
        \sum_{n_3=|n_1-n_2|+1}^{n_1+n_2-1} \phi_{m_3,n_3}\,,
\end{equation}
where the sums are over every other integer. In particular, this
implies that $\phi_3$ has to generate a degenerate representation,
{\it i.e.} that the Verma module generated from $\phi_3$ has a
singular vector. Because of the invariance of the three-point
functions under M\"obius transformations and the symmetry of the
fusion rules (\ref{eq:fusion-rule}) under cyclic permutations, it is
then clear that the three-point function will also satisfy the
condition coming from the null-vector of $\phi_{m_3,n_3}$ at level
$m_3 n_3$.

In general however, this is not sufficient to guarantee that the {\it
irreducible} representation generated from $\phi_3$ possesses a
non-vanishing three-point function with $\phi_1$ and $\phi_2$ --- for
example for integer $t$, if $n_3$ is not in the fundamental domain,
$\phi_{m_3,n_3}$ has typically another singular vector at lower level,
which might then not be null in the three-point function. On the other
hand, it is clear that (\ref{eq:fusion-rule}) gives the correct fusion
rules for the corresponding irreducible representations if $t$ is
irrational, as the degenerate representations possess only one
singular vector in this case. In the notation of the previous section
this means
\begin{equation}\label{eq:fusion-irrat}
\left( {\cal V}_{m_1,n_1} \otimes {\cal V}_{m_2,n_2} \right)_{\mathrm f}
       =\bigoplus_{m_3=|m_1-m_2|+1}^{m_1+m_2-1} 
        \bigoplus_{n_3=|n_1-n_2|+1}^{n_1+n_2-1} {\cal V}_{m_3,n_3}\,.
\hspace*{2cm}  \mbox{($t$ irrational)}
\end{equation}

Before turning to the case of integer $t$, let us mention that this
result gives also rise to a character identity. It is clear from the
previous section that we can define a character for the fusion tensor
product by setting
\begin{equation}
\label{eq:tensor-char}
\chi_{({\cal V}_{m_1,n_1} \otimes {\cal V}_{m_2,n_2})_{\mathrm f}}
     (\tau) =
\lim_{N\rightarrow\infty} \Bigl. 
\tr\Bigr|_{({\cal V}_{m_1,n_1} \otimes 
{\cal V}_{m_2,n_2})_{\mathrm f}^N}
\left(q^{L_0 - c/24}\right) \,,
\end{equation}
where $q=\exp(2 \pi i \tau)$. For irrational $t$ we then obtain
\begin{equation}\label{eq:fusion-char}
  \chi_{({\cal V}_{m_1,n_1} \otimes {\cal V}_{m_2,n_2})_{\mathrm f}}
     (\tau) = 
  \sum_{m_3=|m_1-m_2|+1}^{m_1+m_2-1} 
  \sum_{n_3=|n_1-n_2|+1}^{n_1+n_2-1} \chi_{m_3,n_3}(\tau)\,,
\end{equation}
where the character $\chi_{m,n}(\tau)$ is the character
of ${\cal V}_{m,n}$ which is explicitly given as
\begin{equation}
  \chi_{m,n}(\tau) = \tr_{{\cal V}_{m,n}} q^{L_0-c/24} = 
  \eta(\tau)^{-1} q^{\frac{1-c}{24}} 
  \left( q^{h_{m,n}} - q^{h_{m,-n}} 
  \right)\,,
\end{equation}
and $\eta(\tau)$ is the Dedekind $\eta$ function
\begin{equation}
\eta(\tau) = q^{\frac{1}{24}} \prod_{n=1}^{\infty} (1 - q^n) \,. 
\end{equation}

Let us now turn to analysing the case where $t$ is an integer. In
order to be more specific, let us denote by ${\cal W}_3$ the subspace
of ${\cal M}_{m_3,n_3}$ which is null in the three-point function with
$\phi_1$ and $\phi_2$, and all their descendants. 
Because of the
invariance of the three-point function with respect to the action of
the Virasoro algebra, this subspace has to be a subrepresentation and
therefore is generated from a singular vector in ${\cal M}_{m_3,n_3}$.
If we are in the fundamental domain ($n_3\leq t$), then ${\cal W}_3$
is generated from the singular vector at level $m_3 n_3$, and the
quotient space ${\cal M}_{m_3,n_3} / {\cal W}_3$ is irreducible. On
the other hand, this is not always the case, even if we start with
irreducible representations ($n_1,n_2\leq t$). Indeed, if $t<n_3\leq
2t$, then the embedding diagram of the representation corresponding to
$\phi_{m_3,n_3}$ is
\begin{center}
  \begin{picture}(320,80)(-100,-10)
    \multiput(0,40)(80,0){2}{\vbox to 0pt
      {\vss\hbox to 0pt{\hss$\bullet$\hss}\vss}}
    \put(160,40){\vbox to 0pt
      {\vss\hbox to 0pt{\hss$\times$\hss}\vss}}
    \multiput(0,0)(80,0){3}{\vbox to 0pt
      {\vss\hbox to 0pt{\hss$\bullet$\hss}\vss}}
    \put(240,0){\vbox to 0pt
      {\vss\hbox to 0pt{\hss$\times$\hss}\vss}}
    
    \multiput(5,0)(80,0){2}{\vector(1,0){70}}
    \multiput(165,0)(10,0){6}{\line(1,0){6}}
    \put(225,0){\vector(1,0){10}}
    \put(5,40){\vector(1,0){70}}
    \multiput(85,40)(10,0){6}{\line(1,0){6}}
    \put(145,40){\vector(1,0){10}}
    \put(-100,40){\vbox to 0pt
      {\vss\hbox to 0pt{${\cal V}_{1,n_3}\colon$\hss}\vss}}
    \put(-100,0){\vbox to 0pt
      {\vss\hbox to 0pt{${\cal V}_{m_3,n_3}\colon$\hss}\vss}}
    
    \put(0,50){\hbox to 0pt{\hss$\scriptstyle(1,2t-n_3)$\hss}}
    \put(80,50){\hbox to 0pt{\hss$\scriptstyle(2,n_3-t)$\hss}}
    \put(160,50){\hbox to 0pt{\hss$\scriptstyle(3,2t-n_3)$\hss}}

    \put(0,10){\hbox to 0pt{\hss$\scriptstyle(m_3-1,n_3-t)$\hss}}
    \put(80,10){\hbox to 0pt{\hss$\scriptstyle(m_3,2t-n_3)$\hss}}
    \put(160,10){\hbox to 0pt{\hss$\scriptstyle(m_3+1,n_3-t)$\hss}}
    \put(240,10){\hbox to 0pt{\hss$\scriptstyle(m_3+2,2t-n_3)$\hss}}
  \end{picture}
\end{center}
where in the second line $m_3\geq 2$, and $\times$ denotes the
fundamental null vector (corresponding to the singular vector at level
$m_3 n_3$) which certainly generates a subspace contained in ${\cal
W}_3$. The interesting question is then whether ${\cal W}_3$ is
actually larger, and whether it contains also the subrepresentations
generated from the singular vectors $(2,n_3-t)$, $(m_3,2t-n_3)$ and
$(m_3+1,n_3-t)$.
\smallskip

The above embedding diagram implies that 
the null vector condition for $\phi_{m_3,n_3}$ and $t<n_3<2t$
factorises as
\begin{eqnarray*}
  0 = p_{1,n_3}(\alpha_1,\alpha_2) &=& 
  p_{1,2t-n_3}(\alpha_1,\alpha_2)\, p_{2,n_3-t}(\alpha_1,\alpha_2)\,, \\
  0 = p_{m_3,n_3}(\alpha_1,\alpha_2) &=& 
  p_{m_3-1,n_3-t}(\alpha_1,\alpha_2)\, 
  p_{m_3,2t-n_3}(\alpha_1,\alpha_2)\,
  p_{m_3+1,n_3-t}(\alpha_1,\alpha_2)\,.
\end{eqnarray*}
Furthermore, $t< n_3 \leq n_1+n_2-1$ implies that $2t-n_3 \geq
2t-n_1-n_2+1 \geq |n_1-n_2|+1$; hence, if the right hand side of the
fusion rule (\ref{eq:fusion-rule}) contains a field $\phi_{m_3,n_3}$,
it also contains the field $\phi_{m_3,2t-n_3}$. This implies that
$p_{m_3,2t-n_3}(\alpha_1, \alpha_2)=0$, and thus 
\begin{eqnarray*}
  0 &=& p_{1,2t-n_3}(\alpha_1,\alpha_2), \\
  0 &=& p_{m_3-1,n_3-t}(\alpha_1,\alpha_2) 
  p_{m_3,2t-n_3}(\alpha_1,\alpha_2) = 
  p_{1,m_3t+t-n_3}(\alpha_1,\alpha_2).
\end{eqnarray*}
It follows
that the singular vectors $(2,n_3-t)$ and
$(m_3+1,n_3-t)$ are indeed in ${\cal W}_3$. On the other hand, the
singular vector $(m_3,2t-n_3)$ is not in ${\cal W}_3$, since
the field $\phi_{m_3-1,n_3-t}$ (whose conformal weight equals that of
$\phi_{m_3,n_3}$) is not in the fusion product because of
parity, and thus
\begin{displaymath}
  p_{m_3-1,n_3-t}(\alpha_1,\alpha_2) \neq 0 \,.
\end{displaymath}

Hence we have
\begin{equation}
\label{reducible}
  ({\cal M}_{m_3,n_3}/{\cal W}_3) = \left\{
  \begin{array}{cl}
    {\cal V}_{m_3,n_3} & \hbox{if $n_3\leq t$}\,, \\
   {\cal V}_{1,m_3t+t-n_3} & \hbox{if $n_3> t$}\,.
  \end{array}\right. 
\end{equation}

We have thus shown, that the fusion product has non-vanishing
correlation functions with these reducible subrepresentations. This is
not yet sufficient to determine the fusion product uniquely, as we can
(in this way) only study correlation functions of the fusion product
with highest weight representations (which have in particular a cyclic
highest weight vector). However, as will become apparent from the
explicit calculations, the fusion product typically contains
generalised highest weight representations for which this is not true.
On the other hand, the above analysis certainly imposes constraints on
the possible structure of the fusion product.

\section{The structure of the fusion product}
\label{sec:reps}

We shall now explain what we conjecture to be the structure of the
fusion product. We shall then check that it satisfies the constraints
of the above analysis. In the next section we shall present the
results of the analysis of the fusion product using the algorithm of
Section \ref{sec:fusion}, which confirm these conjectures.

As is clear from the above arguments, the highest weight of the
representation ${\cal M}_{m,n}$ with $2t>n>t$ differs from that of the
representation ${\cal M}_{m,2t-n}$ by an integer. It is then possible
that the two representations couple to form a generalised highest
weight representation. Indeed, this is what seems to happen, and we
conjecture that the decomposition of the fusion product for integer
$t$ is
\begin{equation}
\label{eq:decomposition}
  ({\cal V}_{m_1,n_1} \otimes {\cal V}_{m_2,n_2})_{\mathrm f} 
         = \left\{
  \begin{array}{ll}
    \bigoplus\limits_m
    \bigoplus\limits_{n=|n_1-n_2|+1}^{n_1+n_2-1} 
    {\cal V}_{m,n} & \hbox{if $n_1+n_2\leq t$,} \\[\bigskipamount]
    \bigoplus\limits_m
    \Bigg[ \left(
    \bigoplus\limits_{n=|n_1-n_2|+1}^{2t-n_1-n_2-1} \!\!
    {\cal V}_{m,n} \right) \oplus \left(
    \bigoplus\limits_{n=2t-n_1-n_2+1}^{t-1} \!\!\!\!\!\!\!
    {\cal R}_{m,n} \right) \oplus 
    \underline{{\cal V}_{m,t}} \Bigg] &\hbox{if $n_1+n_2>t$,}
  \end{array}\right.
\end{equation}
where we sum over $m=|m_1-m_2|+1,\ldots,m_1+m_2-3,m_1+m_2-1$ and
$n+n_1+n_2$ odd. The underlined term is only present if
$t+n_1+n_2$ is odd. ${\cal R}_{m,n}$ denotes the decomposable
representation which can schematically be represented by
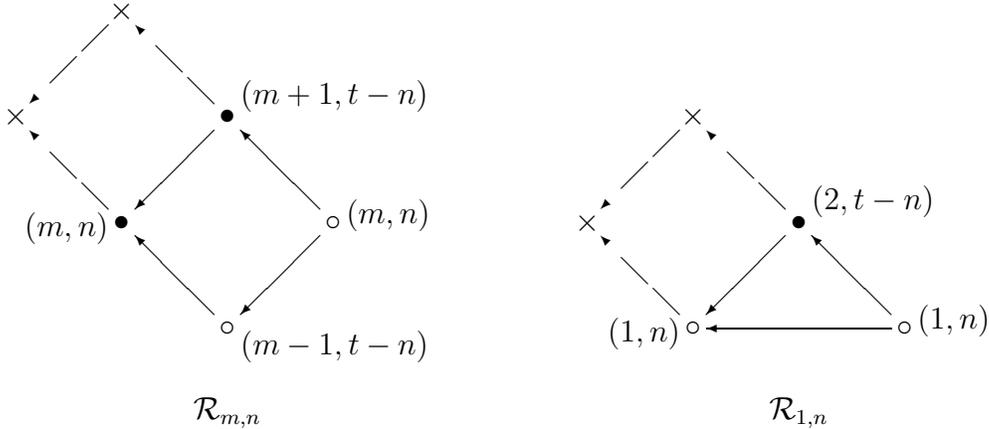
\begin{center}
  \begin{tabular}{c@{\hskip0.5in}c}
  \begin{picture}(180,170)(-10,-20)
    \put(80,0){\vbox to 0pt
      {\vss\hbox to 0pt{\hss$\circ$\hss}\vss}}
    \put(120,40){\vbox to 0pt
      {\vss\hbox to 0pt{\hss$\circ$\hss}\vss}}
    \put(40,40){\vbox to 0pt
      {\vss\hbox to 0pt{\hss$\bullet$\hss}\vss}}
    \put(80,80){\vbox to 0pt
      {\vss\hbox to 0pt{\hss$\bullet$\hss}\vss}}
    \put(0,80){\vbox to 0pt
      {\vss\hbox to 0pt{\hss$\times$\hss}\vss}}
    \put(40,120){\vbox to 0pt
      {\vss\hbox to 0pt{\hss$\times$\hss}\vss}}

    \put(75,5){\vector(-1,1){30}}
    \put(115,35){\vector(-1,-1){30}}
    \put(115,45){\vector(-1,1){30}}
    \put(75,75){\vector(-1,-1){30}}

    \multiput(35,45)(-12,12){2}{\line(-1,1){10}}
    \put(11,69){\vector(-1,1){6}}
    \multiput(75,85)(-12,12){2}{\line(-1,1){10}}
    \put(51,109){\vector(-1,1){6}}
    \multiput(35,115)(-12,-12){2}{\line(-1,-1){10}}
    \put(11,91){\vector(-1,-1){6}}

    \put(85,-10){$(m-1,t-n)$}   
    \put(125,40){$(m,n)$}       
    \put(85,85){$(m+1,t-n)$}
    \put(35,35){\hbox to 0pt{\hss$(m,n)$}}
  \end{picture}
&
  \begin{picture}(180,110)(-10,20)
    \multiput(40,40)(80,0){2}{\vbox to 0pt
      {\vss\hbox to 0pt{\hss$\circ$\hss}\vss}}
    \put(80,80){\vbox to 0pt
      {\vss\hbox to 0pt{\hss$\bullet$\hss}\vss}}
    \put(0,80){\vbox to 0pt
      {\vss\hbox to 0pt{\hss$\times$\hss}\vss}}
    \put(40,120){\vbox to 0pt
      {\vss\hbox to 0pt{\hss$\times$\hss}\vss}}

    \put(115,40){\vector(-1,0){70}}
    \put(115,45){\vector(-1,1){30}}
    \put(75,75){\vector(-1,-1){30}}

    \multiput(35,45)(-12,12){2}{\line(-1,1){10}}
    \put(11,69){\vector(-1,1){6}}
    \multiput(75,85)(-12,12){2}{\line(-1,1){10}}
    \put(51,109){\vector(-1,1){6}}
    \multiput(35,115)(-12,-12){2}{\line(-1,-1){10}}
    \put(11,91){\vector(-1,-1){6}}

    \put(125,40){$(1,n)$}       
    \put(85,85){$(2,t-n)$}
    \put(35,35){\hbox to 0pt{\hss$(1,n)$}}
  \end{picture}
\\
  ${\cal R}_{m,n}$ & ${\cal R}_{1,n}$
  \end{tabular}
\end{center}
Here, each vertex $\bullet$ or $\circ$ with label $(r,s)$ denotes a
state of conformal weight $h_{r,s}$ and the vertices $\times$
correspond to null vectors. An arrow $A\longrightarrow B$ indicates
that the vertex $B$ is in the image of $A$ under the action of the
Virasoro algebra. Conformal weights are constant horizontally, and
increase vertically. The vertices $\circ$ denote states which are not
descendents, {\it i.e.}\ states which cannot be obtained by the action
of the negative modes from other states.

We should mention that the indecomposable representations 
${\cal R}_{m,n}$ with $m\geq 2$ appear to be decomposable if
only the lowest energy space of the product is analysed in the way
proposed by Nahm \cite{Nahm94}; to see that they are in fact
indecomposable it is necessary to analyse the product space up to
level $(m-1) (t-n)$. 
\medskip

Before describing these representations in more detail, let us first
explain, why this is consistent with the above analysis. Because of
M\"obius covariance we can consider the case where the highest weight
representation is inserted at infinity. Then, by a standard argument
of conformal field theory, it is clear that any singular vector in
the highest weight representation at infinity can only get a
non-vanishing contribution from states in the fusion product which
are not descendents; these states have been denoted by $\circ$ in the
above diagram.

Let us now consider the three point function 
$\langle \phi_3 \phi_2 \phi_1\rangle$, and in particular the
contribution of the subrepresentation ${\cal R}_{m,n}$ in the fusion
product of $\phi_1$ and $\phi_2$. For $m\geq 2$, the first singular
vector $(m_3+1,t-n_3)$ of the highest weight representation
$(m_3,n_3)=(m,n)$ (where $0<n<t$) is null in the correlation function
with the representation ${\cal R}_{m,n}$, as there does not exist a
non-descendent state in ${\cal R}_{m,n}$ at this level. The same is
also true for the second singular vector $(m_3+1,n_3-t)=(m+1,t-n)$ 
of the highest weight representation $(m_3,n_3)=(m,2t-n)$. On the
other hand, the first singular vector $(m_3,2t-n_3)=(m,n)$ of the
highest weight representation $(m_3,n_3)=(m,2t-n)$ is not null (with
respect to ${\cal R}_{m,n}$), as it gets a non-vanishing 
contribution from the  $\circ$-state $(m,n)$ in ${\cal R}_{m,n}$. 

For $m=m_3=1$, all singular vectors of the two highest weight
representations at infinity vanish in correlation functions with 
${\cal R}_{m,n}$, as all states in ${\cal R}_{m,n}$ of the
corresponding level are descendents. Thus the conjecture is consistent
with the above findings. 
\smallskip 

The conjecture also satisfies another constraint which comes from the
analysis of the characters. We can easily read off the 
character of ${\cal R}_{m,n}$ from the diagram, and find
\begin{eqnarray}
  \chi_{{\cal R}_{m,n}}(\tau) &=& 
  \chi_{{\cal V}_{m,n}}(\tau) + \chi_{{\cal V}_{m,2t-n}}(\tau) \\*
  &=& \left\{
  \begin{array}{ll}
    2 \chi_{1,n}(\tau) + \chi_{2,t-n}(\tau) & \hbox{for $m=1$} \\
    \chi_{m-1,t-n}(\tau) + 2 \chi_{m,n}(\tau) + \chi_{m+1,t-n}(\tau) &
    \hbox{for $m>1$} 
  \end{array}\right.
\end{eqnarray}
In particular, this implies that the character of the fusion product
satisfies (\ref{eq:fusion-char}), even for integer $t$. This is in
accordance with the observation that the structure of the character
(\ref{eq:tensor-char}) should be unchanged when irrational $t$
approaches an integer, as the structure of the representations ${\cal
V}_{m_1,n_1}$ and ${\cal V}_{m_2,n_2}$ is unchanged in this limit: the
additional vectors which become singular as $t$ becomes an integer are
already in the subspace generated by the null vector at level $m_1
n_1$ and $m_2 n_2$, respectively.
\smallskip

We close this section by describing the structure of the
generalised highest weight representations ${\cal R}_{m,n}$ in more
detail. In the following diagram we have denoted actual states of
${\cal R}_{m,n}$ by $\bullet$, and vectors in the Verma module which are
null in ${\cal R}_{m,n}$ by $\times$ if they are singular, and by
$\boxtimes$ if they are sub-singular. (The distinction
between singular and sub-singular will be explained later.) 
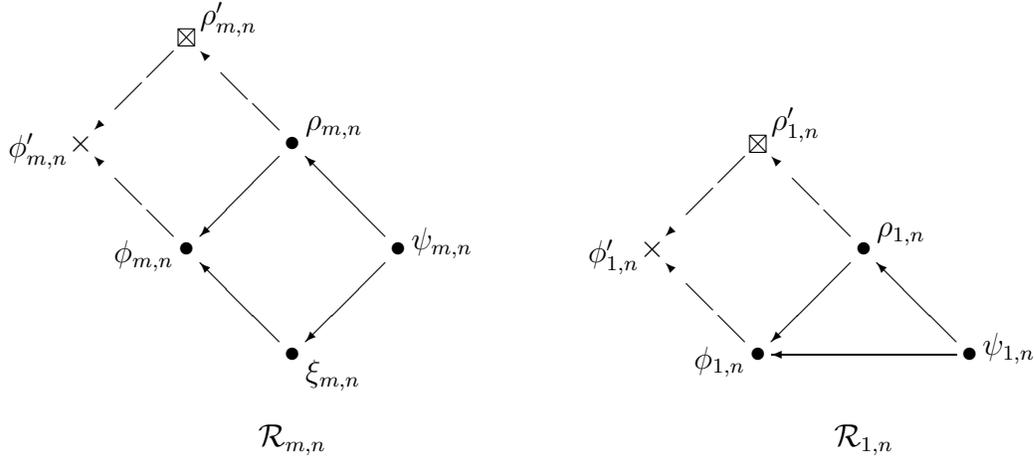
\begin{center}
  \begin{tabular}{c@{\hskip0.5in}c}
  \begin{picture}(180,180)(-10,-20)
    \put(80,0){\vbox to 0pt
      {\vss\hbox to 0pt{\hss$\bullet$\hss}\vss}}
    \multiput(40,40)(80,0){2}{\vbox to 0pt
      {\vss\hbox to 0pt{\hss$\bullet$\hss}\vss}}
    \put(80,80){\vbox to 0pt
      {\vss\hbox to 0pt{\hss$\bullet$\hss}\vss}}
    \put(0,80){\vbox to 0pt
      {\vss\hbox to 0pt{\hss$\times$\hss}\vss}}
    \put(40,120){\vbox to 0pt
      {\vss\hbox to 0pt{\hss$\boxtimes$\hss}\vss}}

    \put(75,5){\vector(-1,1){30}}
    \put(115,35){\vector(-1,-1){30}}
    \put(115,45){\vector(-1,1){30}}
    \put(75,75){\vector(-1,-1){30}}

    \multiput(35,45)(-12,12){2}{\line(-1,1){10}}
    \put(11,69){\vector(-1,1){6}}
    \multiput(75,85)(-12,12){2}{\line(-1,1){10}}
    \put(51,109){\vector(-1,1){6}}
    \multiput(35,115)(-12,-12){2}{\line(-1,-1){10}}
    \put(11,91){\vector(-1,-1){6}}

    \put(85,-10){$\xi_{m,n}$}
    \put(125,40){$\psi_{m,n}$}
    \put(85,85){$\rho_{m,n}$}
    \put(35,35){\hbox to 0pt{\hss$\phi_{m,n}$}}
    \put(-5,75){\hbox to 0pt{\hss$\phi'_{m,n}$}}
    \put(45,125){$\rho'_{m,n}$}
  \end{picture}
&
  \begin{picture}(180,120)(-10,20)
    \multiput(40,40)(80,0){2}{\vbox to 0pt
      {\vss\hbox to 0pt{\hss$\bullet$\hss}\vss}}
    \put(80,80){\vbox to 0pt
      {\vss\hbox to 0pt{\hss$\bullet$\hss}\vss}}
    \put(0,80){\vbox to 0pt
      {\vss\hbox to 0pt{\hss$\times$\hss}\vss}}
    \put(40,120){\vbox to 0pt
      {\vss\hbox to 0pt{\hss$\boxtimes$\hss}\vss}}

    \put(115,40){\vector(-1,0){70}}
    \put(115,45){\vector(-1,1){30}}
    \put(75,75){\vector(-1,-1){30}}

    \multiput(35,45)(-12,12){2}{\line(-1,1){10}}
    \put(11,69){\vector(-1,1){6}}
    \multiput(75,85)(-12,12){2}{\line(-1,1){10}}
    \put(51,109){\vector(-1,1){6}}
    \multiput(35,115)(-12,-12){2}{\line(-1,-1){10}}
    \put(11,91){\vector(-1,-1){6}}

    \put(125,40){$\psi_{1,n}$}
    \put(85,85){$\rho_{1,n}$}
    \put(35,35){\hbox to 0pt{\hss$\phi_{1,n}$}}
    \put(-5,75){\hbox to 0pt{\hss$\phi'_{1,n}$}}
    \put(45,125){$\rho'_{1,n}$}
  \end{picture}
\\
  ${\cal R}_{m,n}$ & ${\cal R}_{1,n}$
  \end{tabular}
\end{center}

The representation ${\cal R}_{m,n}$ contains a subrepresentation
${\cal V}_{1,(m-1)t+n}$ which is generated from a highest weight
vector $\xi_{m,n}$ of weight $h_{m-1,t-n}$, {\it i.e.}
\begin{eqnarray*}
    L_0 \xi_{m,n} &=& h_{m-1,t-n} \xi_{m,n}\,, \\*
    L_{p} \xi_{m,n} &=& 0\,, \qquad\hbox{for $p\geq 1$}\,.
\end{eqnarray*}
Unless $m=1$, the representation ${\cal V}_{1,(m-1)t+n}$ is reducible
and we denote by $\phi_{m,n}$ its lowest singular vector,
\begin{eqnarray*} 
  \sigma_{m-1,t-n} \xi_{m,n} &=& \phi_{m,n}\,.
\end{eqnarray*}
Here $\sigma_{p,q}$ denotes the monomial of negative Virasoro modes
which generates the singular vector at level $p q$ from the highest weight
vector; we have fixed the normalisation of $\sigma_{p,q}$ by requiring
that the coefficient of $L_{-1}^{p q}$ is $1$.

The highest weight representation ${\cal V}_{m,n}$ generated from
$\phi_{m,n}$ is an irreducible subrepresentation of ${\cal R}_{m,n}$,
\begin{eqnarray*} 
L_0 \phi_{m,n} &=& h_{m,n} \phi_{m,n}\,, \\
\sigma_{m,n} \phi_{m,n} &=& \phi'_{m,n} \equiv 0\,.  
\end{eqnarray*}
The total representation ${\cal R}_{m,n}$ is generated from a cyclic
vector $\psi_{m,n}$ which is not an eigenvector of $L_0$ but forms a
Jordan cell of dimension two with the vector $\phi_{m,n}$ of conformal
weight $h_{m,n}$,
\begin{eqnarray*} 
L_0 \psi_{m,n} &=& h_{m,n} \psi_{m,n} + \phi_{m,n}\,, \\ 
L_0 \phi_{m,n} &=& h_{m,n} \phi_{m,n}\,.  
\end{eqnarray*}
Unless $m=1$, $\psi_{m,n}$ is not a highest weight vector, but
positive Virasoro modes map $\psi_{m,n}$ to descendants of
$\xi_{m,n}$. We can redefine $\psi_{m,n}$ (by adding descendants of
$\xi_{m,n}$) so that it is annihilated by $L_p$ with $p\geq
2$. (This is possible because there is only one singular
vector in the space of descendants of $\xi_{m,n}$ at level
$(m-1)(t-n)$.) This fixes the freedom to define $\psi_{m,n}$ up to
the addition of multiples of $\phi_{m,n}$. $L_1 \psi_{m,n}$ (which does
not depend on this freedom) is then annihilated by $L_p$ with $p\geq
2$, and since there is only one such state at level $(m-1)(t-n) - 1$
in the representation generated from $\xi_{m,n}$ (as the inner product
is non-degenerate), it is unique up to a factor.\footnote{We thank
Falk Rohsiepe for pointing out to us that there is only one
characteristic parameter in this case. See also 
\cite{Roh96} for a more detailed analysis of these
representations.} It follows that 
$ L_1^{(m-1)(t-n)} \psi_{m,n}$ does not vanish (if $L_1 \psi_{m,n}\neq
0$), and we can thus  describe the remaining factor by
\begin{equation}
  \begin{array}{rcl}
    L_1^{(m-1)(t-n)} \psi_{m,n} &=& \beta_{m,n} \; \xi_{m,n}\, \\
    L_{p} \psi_{m,n} &=& 0\,, \qquad\hbox{for $p \geq 2$}\,.
  \end{array}
  \label{eq:psi-choice}
\end{equation}
Different values for $\beta_{m,n}$ give rise to inequivalent
representations. In particular, we cannot scale $\beta_{m,n}$ away by
redefining $\psi_{m,n}$ or $\xi_{m,n}$, as we have the additional
relation 
\begin{equation}
\left( L_0 - h_{m,n}\right) \psi_{m,n} = \sigma_{m-1,t-n} \xi_{m,n} \,.
\end{equation}
For the specific representations obtained in the fusion product of
known representations these parameters are determined. We have
calculated $\beta_{m,n}$ in some cases, and have included the explicit
results in the next section.  

The Verma module corresponding to ${\cal R}_{m,n}$, {\it i.e.} the
module which is generated by the action of the negative modes on 
$\xi_{m,n}$ and $\psi_{m,n}$, contains a
sub-singular vector $\rho_{m,n}$ of weight $h_{m+1,t-n}$. This means
that $\rho_{m,n}$ is not singular in the Verma module itself, but that
it becomes singular when we quotient the Verma module by the
subrepresentation generated from $\phi_{m,n}$. (Another way of saying
this is that the positive modes map $\rho_{m,n}$ to descendants of
$\phi_{m,n}$.) $\rho_{m,n}$ is not null in ${\cal R}_{m,n}$, since the
singular vector $\phi_{m,n}$ is not null either. We normalise $\rho_{m,n}$
by requiring that it is of the form
\begin{equation}
  \rho_{m,n} = \sigma_{m,n} \psi_{m,n} + L_{-P} \xi_{m,n}\,,
\end{equation}
where $|P|=(m-1) t + n$, and $\sigma_{p,q}$ is as before. 

The next sub-singular vector in the Verma module is of weight 
$h_{m+2,n}$ and is denoted by $\rho'_{m,n}$. It is actually null in 
${\cal R}_{m,n}$, as the positive modes map 
$\rho'_{m,n}$ to descendants of $\phi'_{m,n}$ which is also null.
Again, we can assume that $\rho'_{m,n}$ is of the form 
\begin{equation}
  \rho'_{m,n} = \sigma_{m+1,t-n} \rho_{m,n} + L_{-Q} \phi_{m,n}\,,
\end{equation}
where $|Q|=(m+1) t - n$.

For $m=1$, the situation is degenerate, as $\xi_{1,n} = 0$, and
$\psi_{1,n}$ is annihilated by all positive modes,
\begin{equation} 
L_{n} \psi_{m,n} = 0\,, \qquad\hbox{for $n>0$}\,.
\end{equation}
In particular, this means that all representations of this type are
equivalent, and that in this case there is no free parameter labelling
inequivalent representations. 

We can also explicitly give a basis for ${\cal R}_{m,n}^N$, the
subspace of ${\cal R}_{m,n}$ of all descendents up to level $N$; it
can be taken to be the following subset of the lexicographically
ordered states
\begin{equation}
\label{basis}
  \begin{array}{rll}
   {\displaystyle \cdots L_{-2}^{k_2} L_{-1}^{k_1} \xi_{m,n}\,, }
    & 
   {\displaystyle \sum_j jk_j \leq N\,,}
    & 
   {\displaystyle k_1<(m-1)(t-n)\,,} \\
   {\displaystyle \cdots L_{-2}^{k'_2} L_{-1}^{k'_1} \phi_{m,n}\,,} 
    & 
   {\displaystyle \sum_j jk_j \leq N - (m-1)(t-n)\,,}
    & 
   {\displaystyle k_1<mn\,,} \\
   {\displaystyle \cdots L_{-2}^{k'_2} L_{-1}^{k'_1} \psi_{m,n}\,,} 
    & 
   {\displaystyle \sum_j jk_j \leq N\,,}
    & 
   {\displaystyle k_1<mn\,,} \\
   {\displaystyle \cdots L_{-2}^{k''_2} L_{-1}^{k''_1} \rho_{m,n}\,,} 
    & 
   {\displaystyle \sum_j jk_j \leq N - mn\,,}
    &
   {\displaystyle k_1<(m+1)(t-n)\,.} 
  \end{array}
\end{equation}
\medskip

\section{Explicit Calculations}
\label{sec:results}

In this section we describe the explicit calculations which we have
done in order to determine the fusion product of some of the
representations.  We shall outline first how we proceeded in general,
before giving our results in detail. For one example we have included
the explicit matrices, describing the action of the positive Virasoro
generators on the fusion product, in the appendix.

Let us suppose that ${\cal V}_1$ and ${\cal V}_2$ are two
quasirational representations, and that we want to analyse their
fusion product up to a given level $L$.\footnote{For most of the
  following we chose $L=6$.} 
We consider first the space ${\cal F}_{12} = {\cal V}_1^{\mathrm{s}}
\otimes {\cal V}_{2}^L$, which contains the space we are interested
in, $({\cal V}_1 \otimes {\cal V}_{2})_{\mathrm{f}}^L$, by the
arguments of Section \ref{sec:fusion}.  Typically ${\cal F}_{12}$ is
too large, reflecting that there exists a `spurious subspace'
\cite{Nahm94}. To find the additional relations we calculate
$\Delta(a) \Psi$, where $\Psi\in{\cal F}_{12}$ and $a\in{\cal
  A}_{L+1}$, using the comultiplication. We then use the
null-relations in ${\cal V}_1$ and ${\cal V}_2$ to rewrite the result,
and apply then the reduction algorithm of Section \ref{sec:fusion} to
obtain an expression in ${\cal F}_{12}$. By construction we know that
this expression is in the subspace by which we have to quotient
$({\cal V}_1 \otimes {\cal V}_{2})_{\mathrm{f}}$ in order to obtain
$({\cal V}_1 \otimes {\cal V}_{2})^L_{\mathrm{f}}$; it may, however,
happen that the resulting expression is not identically zero, and this
gives then rise to a relation in ${\cal F}_{12}$.

A priori, it is not clear how to find all missing relations. We took
$a$ to be all monomials of a fixed weight greater than $L$, and
$\Psi\in {\cal V}_1^{\mathrm{s}} \otimes {\cal V}_2^0$.  In most cases
it was sufficient to consider monomials of weight $L+1$ but in some
cases we had to increase successively the weight of $a$ up to $L+5$ to
reduce the space ${\cal F}_{12}$ to the dimension conjectured in the
previous section.  We also checked (in some cases) that no further
relations arose when we did the calculation for some monomials $a$ of
higher weight.

Having obtained the space $({\cal V}_1\otimes{\cal
  V}_2)_{\mathrm{f}}^L$ we construct the spaces $({\cal
  V}_1\otimes{\cal V}_2)_{\mathrm{f}}^n$ for $0\leq n<L$ by
successively imposing the constraints $\Delta(a)=0$ for all $a$ of
weight $n+1$. The Virasoro generators map
\begin{displaymath}
  L_m\colon ({\cal V}_1\otimes{\cal V}_2)_{\mathrm{f}}^n \to 
  ({\cal V}_1\otimes{\cal V}_2)_{\mathrm{f}}^{n-m} \,,
\end{displaymath}
and we can thus calculate the matrices, representing $L_m$ in some
basis. At first, these bases are chosen at random by the computer
program, and in order to interpret the results more easily, we change
to a basis in which the action is in block-diagonal form.  It is then
easy to check the conjectures about the decomposition of the fusion
product. To read off the parameter characterising the generalised
highest weight representations ${\cal R}_{m,n}$, we change the basis
in the corresponding block to the canonical basis introduced in the
previous section.

\subsection{$t=2$}
\label{sec:r2}

We first analysed the fusion product of irreducible
representations. Up to level six the decomposition was as follows
\begin{eqnarray*}
  \left({\cal V}_{2,1}\otimes{\cal V}_{2,1}\right)_{\mathrm f} &=& 
    {\cal V}_{1,1} \oplus {\cal V}_{3,1} \,, \\
  \left({\cal V}_{m,1}\otimes{\cal V}_{1,2}\right)_{\mathrm f} &=& 
    {\cal V}_{m,2}\,, \qquad\hbox{for $m=1, \ldots, 5$}\,, \\
  \left( {\cal V}_{1,2}\otimes{\cal V}_{m,2} \right)_{\mathrm f} &=&
  {\cal R}_{m,1}\,, \qquad\hbox{for $m=1,\ldots,5$}\,, \\
  \left( {\cal V}_{2,2}\otimes{\cal V}_{2,2} \right)_{\mathrm f} &=&
  {\cal R}_{1,1} \oplus {\cal R}_{3,1}\,, \\
  \left( {\cal V}_{2,2}\otimes{\cal V}_{3,2} \right)_{\mathrm f} &=&
  {\cal R}_{2,1} \oplus {\cal R}_{4,1}\,, \\
  \left( {\cal V}_{3,2}\otimes{\cal V}_{3,2} \right)_{\mathrm f} &=&
  {\cal R}_{1,1} \oplus {\cal R}_{3,1} \oplus {\cal R}_{5,1}\,.
\end{eqnarray*}
We should note that the fusion of $\left({\cal V}_{1,2}\otimes{\cal
V}_{m,2}\right)$ provides a way of constructing the generalised
highest weight representation ${\cal R}_{m,1}$. We have used this to
read off the characteristic parameters of ${\cal R}_{m,1}$, and the
results are contained in Table \ref{tab:data2}. For the case of 
$({\cal V}_{2,1}\otimes{\cal V}_{1,2})_{\mathrm f} = {\cal R}_{2,1}$
we have included some more details in the appendix.

\begin{table}[bht]
\begin{displaymath}
\renewcommand{\arraystretch}{1.5}
  \begin{array}{cccc}
    \hline
    (m,n) & h_{m,n} & h_{m-1,t-n} & \beta_{m,n} \\\hline
    (1,1) & 0 & 0 & -\\
    (2,1) & 1 & 0 & -1 \\
    (3,1) & 3 & 1 & -18 \\
    (4,1) & 6 & 3 & -2700 \\
    (5,1) & 10 & 6 & -1587600 \\\hline
  \end{array}
\end{displaymath}
  \caption{Representation data of ${\cal R}_{m,1}$ for $t=2$.}
  \label{tab:data2}
\end{table}

We have checked that all representations of type ${\cal R}_{m,n}$
which appear in the above decompositions are indeed equivalent
representations, {\it i.e.} have the same characteristic parameter
(see Table \ref{tab:data2}). This provides a partial consistency check of
our results, as the associativity and symmetry of the fusion product
imply certain restrictions. For example, as ${\cal V}_{1,1}$ is the
identity field, we find
\begin{eqnarray*}
\left( {\cal V}_{2,2}\otimes{\cal V}_{2,2} \right)_{\mathrm f} &=&
\left( {\cal V}_{2,1}\otimes{\cal V}_{1,2}\otimes{\cal V}_{2,1} 
\otimes{\cal V}_{1,2} \right)_{\mathrm f} \\*
& = & \Bigl({\cal R}_{1,1} \otimes \left( {\cal V}_{1,1}\oplus
{\cal V}_{3,1}\right)\Bigr)_{\mathrm f} \\*
& = & {\cal R}_{1,1} \oplus \left({\cal R}_{1,1} \otimes 
{\cal V}_{3,1}\right)_{\mathrm f}\,.
\end{eqnarray*}

We should note that the Verma module corresponding to ${\cal R}_{1,1}$
contains a subrepresentation of the same structure as ${\cal R}_{2,1}$. 
Indeed, we have
\begin{equation}
\begin{array}{rcl}
\psi_{2,1} &\cong& \rho_{1,1} = L_{-1} \psi_{1,1} \\
\phi_{2,1} &\cong& \phi'_{1,1} = L_{-1} \phi_{1,1} \\
\xi_{2,1} &\cong& \phi_{1,1} = \frac{1}{2} L_{1} \rho_{1,1}\,.
\end{array}
\end{equation}
However, the last line implies that its characteristic parameter
($\frac{1}{2}$) is different from that of ${\cal R}_{2,1}$ in Table
\ref{tab:data2} ($-1$).  
\smallskip

The results suggest that we have for general $m_1, m_2$, 
\begin{equation}
\label{eq:asume1}
  \left({\cal V}_{m_1,1}\otimes{\cal V}_{m_2,1}\right)_{\mathrm f} = 
    \bigoplus_{m=|m_1-m_2|+1}^{m_1+m_2-1} {\cal V}_{m,1} \,,
\end{equation}
and that for general $m$
\begin{equation}
\label{eq:assume2}
\begin{array}{rcl}
  \left({\cal V}_{m,1}\otimes{\cal V}_{1,2}\right)_{\mathrm f} & = &
    {\cal V}_{m,2}\,, \\
  \left({\cal V}_{m,2}\otimes{\cal V}_{1,2}\right)_{\mathrm f} & = &
    {\cal R}_{m,1}\,,
\end{array}
\end{equation}
where ${\cal R}_{m,1}$ is a generalised highest weight representation
of the type discussed before, whose characteristic parameter is
described (for $m\leq 5$) by Table \ref{tab:data2}. This confirms the
conjecture of the previous section. 
\smallskip

In a next step we analysed the fusion involving the representation
${\cal R}_{1,1}$,
\begin{equation}
  \begin{array}{rcl}
    \left( {\cal V}_{1,2}\otimes{\cal R}_{1,1} \right)_{\mathrm f} &=& 
    {\cal V}_{1,2} \oplus {\cal V}_{1,2} \oplus {\cal V}_{2,2}\,, \\
    \left( {\cal R}_{1,1}\otimes{\cal R}_{1,1} \right)_{\mathrm f} &=& 
    {\cal R}_{1,1} \oplus {\cal R}_{1,1} \oplus {\cal R}_{2,1}\,.
  \end{array}
  \label{eq:r11fusion}
\end{equation}

Assuming that (\ref{eq:asume1}) and (\ref{eq:assume2}) hold, this is
then already sufficient to derive the decomposition of all fusion
products, using the associativity and symmetry of the fusion
product.\footnote{In fact, the second identity of (\ref{eq:r11fusion})
  follows already from the first, using the associativity of the
  fusion product and (\ref{eq:assume2})} We find that
\begin{eqnarray*}
  \left( {\cal V}_{m_1,1}\otimes{\cal V}_{m_2,2} \right)_{\mathrm f} &=& 
  \left( {\cal V}_{m_1,1}\otimes{\cal V}_{m_2,1}
    \otimes{\cal V}_{1,2} \right)_{\mathrm f} \\*
  &=& \bigoplus_m \left(
    {\cal V}_{m,1}  \otimes{\cal V}_{1,2} 
           \right)_{\mathrm f} \\*
  &=& \bigoplus_m {\cal V}_{m,2}\,, 
\nonumber
\end{eqnarray*}
where (as always in the following) we sum over 
$m=|m_1-m_2|+1, |m_1-m_2|+3, \ldots,m_1+m_2-1$. Similarly
\begin{equation}
\begin{array}{rcl}
  {\displaystyle 
     \left( {\cal V}_{m_1,2}\otimes{\cal V}_{m_2,2} \right)_{\mathrm f}}
  &=& \displaystyle \bigoplus_m {\cal R}_{m,1}\,, \\
  \displaystyle
     \left( {\cal V}_{m_1,1}\otimes{\cal R}_{m_2,1} \right)_{\mathrm f} 
  &=& \displaystyle \bigoplus_m {\cal R}_{m,1}\,, \\
  \displaystyle 
     \left( {\cal V}_{m_1,2}\otimes{\cal R}_{m_2,1} \right)_{\mathrm f} 
  &=& \displaystyle
        \bigoplus_m \left( {\cal V}_{m,2}\oplus{\cal V}_{m,2}\oplus
    \underline{{\cal V}_{m-1,2}}\oplus{\cal V}_{m+1,2} \right)\,, \\
  \displaystyle 
    \left( {\cal R}_{m_1,1}\otimes{\cal R}_{m_2,1} \right)_{\mathrm f}
  &=& \displaystyle
       \bigoplus_m \left( {\cal R}_{m,1}\oplus{\cal R}_{m,1}\oplus
    \underline{{\cal R}_{m-1,1}}\oplus{\cal R}_{m+1,1} \right)\,,
\end{array}
\end{equation}
where the underlined terms on the right hand side are absent for
$m=1$. This demonstrates in particular that the set of representations
${\cal R}_{m,1}$, ${\cal V}_{m,1}$ and ${\cal V}_{m,2}$ is closed
under fusion.

\subsection{$t=3$}
\label{sec:r3}

We calculated the following fusion products. The calculations were
done again up to level six, apart from the last three identities which
were only checked up to level five.  
We note that the last six identities follow from the previous ones by
the associativity and commutativity of the fusion product, and thus
provide a consistency check.
\begin{eqnarray*}
  \left({\cal V}_{2,1}\otimes{\cal V}_{2,1}\right)_{\mathrm f} &=& 
    {\cal V}_{1,1} \oplus {\cal V}_{3,1} \,, \\
  \left({\cal V}_{m,1}\otimes{\cal V}_{1,2}\right)_{\mathrm f} &=& 
    {\cal V}_{m,2}\,, \qquad\hbox{for $m=1, \ldots, 5$}\,, \\
  \left({\cal V}_{m,1}\otimes{\cal V}_{1,3}\right)_{\mathrm f} &=& 
    {\cal V}_{m,3}\,, \qquad\hbox{for $m=1, \ldots, 5$}\,, \\[\medskipamount]
  \left({\cal V}_{1,2}\otimes{\cal V}_{1,2}\right)_{\mathrm f} &=&
    {\cal V}_{1,1} \oplus {\cal V}_{1,3}\,, \\
  \left({\cal V}_{1,2}\otimes{\cal V}_{1,3}\right)_{\mathrm f} &=&
    {\cal R}_{1,2}\,, \\
  \left({\cal V}_{1,3}\otimes{\cal V}_{1,3}\right)_{\mathrm f} &=&
    {\cal R}_{1,1} \oplus {\cal V}_{1,3}\,, \\
  \left({\cal V}_{2,2}\otimes{\cal V}_{1,3}\right)_{\mathrm f} &=&
    {\cal R}_{2,2}\,, \\
  \left({\cal V}_{2,3}\otimes{\cal V}_{1,3}\right)_{\mathrm f} &=&
    {\cal R}_{2,1} \oplus {\cal V}_{2,3}\,, \\[\medskipamount]
  \left({\cal V}_{1,2}\otimes{\cal R}_{1,1}\right)_{\mathrm f} &=& 
    {\cal R}_{1,2} \oplus {\cal V}_{2,3}\,, \\
  \left({\cal V}_{1,2}\otimes{\cal R}_{1,2}\right)_{\mathrm f} &=& 
    {\cal R}_{1,1} \oplus 2{\cal V}_{1,3}\,, \\
  \left({\cal V}_{1,3}\otimes{\cal R}_{1,1}\right)_{\mathrm f} &=& 
    {\cal R}_{2,2} \oplus 2{\cal V}_{1,3}\,, \\
  \left({\cal V}_{1,3}\otimes{\cal R}_{1,2}\right)_{\mathrm f} &=& 
    2{\cal R}_{1,2} \oplus {\cal V}_{2,3}\,, 
    \\[\medskipamount]
  \left({\cal R}_{1,1}\otimes{\cal R}_{1,1}\right)_{\mathrm f} &=& 
    2 {\cal R}_{1,1} \oplus {\cal R}_{2,2} \oplus {\cal V}_{1,3}
    \oplus {\cal V}_{3,3}\,, \\
  \left({\cal R}_{1,1}\otimes{\cal R}_{1,2}\right)_{\mathrm f} &=& 
    2 {\cal R}_{1,2} \oplus {\cal R}_{2,1} \oplus 2{\cal V}_{2,3}\,, \\
  \left({\cal R}_{1,2}\otimes{\cal R}_{1,2}\right)_{\mathrm f} &=& 
    2 {\cal R}_{1,1} \oplus {\cal R}_{2,2} \oplus 4{\cal V}_{1,3}\,.
\end{eqnarray*}
We have included the parameters, characterising the generalised
highest weight representations, in Table \ref{tab:data3}. 
\begin{table}[htb]
\begin{displaymath}
\renewcommand{\arraystretch}{1.5}
  \begin{array}{cccc}
    \hline
    (m,n) & h_{m,n} & h_{m-1,t-n} & \beta_{m,n} \\\hline
    (1,1) & 0 & 0 & - \\
    (1,2) & -\frac14 & -\frac14 & - \\
    (2,1) & \frac74 & -\frac14 & \frac89 \\
    (2,2) & 1 & 0 & -2
    \\\hline
  \end{array}
\end{displaymath}
  \caption{Representation data of ${\cal R}_{m,n}$ for $t=3$.}
  \label{tab:data3}
\end{table}
Again, as before, we checked explicitly that all the
representations denoted by ${\cal R}_{m,n}$ are indeed equivalent
representations, {\it i.e.} have the same characteristic parameter.

The results suggest that we have for all $m_1$ and $m_2$
\begin{equation}
\label{eq:assume3}
  \left({\cal V}_{m_1,1}\otimes{\cal V}_{m_2,1}\right)_{\mathrm f} = 
    \bigoplus_{m=|m_1-m_2|+1}^{m_1+m_2-1} {\cal V}_{m,1} \,,
\end{equation}
and that for all $m$
\begin{equation}
\label{eq:assume4}
\begin{array}{rcl}
  \left({\cal V}_{m,1}\otimes{\cal V}_{1,2}\right)_{\mathrm f} &=& 
    {\cal V}_{m,2}\,, \\
  \left({\cal V}_{m,1}\otimes{\cal V}_{1,3}\right)_{\mathrm f} &=& 
    {\cal V}_{m,3}\,, \\
  \left({\cal V}_{m,2}\otimes{\cal V}_{1,3}\right)_{\mathrm f} &=& 
    {\cal R}_{m,2}\,, \\
  \left({\cal V}_{m,3}\otimes{\cal V}_{1,3}\right)_{\mathrm f} &=& 
    {\cal V}_{m,3} \oplus {\cal R}_{m,1}\,,
\end{array}
\end{equation}
where ${\cal R}_{m,n}$ is a generalised highest weight representation
of the type discussed before, whose characteristic parameter is
described by Table \ref{tab:data3}. This again confirms our
conjectures of the previous section. 

Together with the above results about the fusion of the generalised
highest weight representations, (\ref{eq:assume3}) and
(\ref{eq:assume4}) are then sufficient to derive the decomposition of
all fusion products for $t=3$, using the associativity and symmetry of
the fusion product. We find
\begin{eqnarray}
  \left({\cal V}_{m_1,n_1}\otimes{\cal V}_{m_2,n_2}\right)_{\mathrm f} &=& 
    \bigoplus_{m=|m_1-m_2|+1}^{m_1+m_2-1} \left(
      \bigoplus_{n=|n_1-n_2|+1}^{n_1+n_2-1}{\cal V}_{m,n} \right)\,,
\end{eqnarray}
where $(n_1,n_2)=(1,1); (1,2); (1,3)$ or $(2,2)$. Furthermore
\begin{eqnarray*}
  \left({\cal V}_{m_1,2}\otimes{\cal V}_{m_2,3}\right)_{\mathrm f} &=& 
    \bigoplus_m {\cal R}_{m,2} \,, \\
  \left({\cal V}_{m_1,3}\otimes{\cal V}_{m_2,3}\right)_{\mathrm f} &=& 
    \bigoplus_m \left({\cal V}_{m,3} \oplus {\cal R}_{m,1}\right)\,,
    \\[\medskipamount] 
  \left({\cal V}_{m_1,1}\otimes{\cal R}_{m_2,1}\right)_{\mathrm f} &=& 
    \bigoplus_m{\cal R}_{m,1}\,, \\
  \left({\cal V}_{m_1,1}\otimes{\cal R}_{m_2,2}\right)_{\mathrm f} &=& 
    \bigoplus_m{\cal R}_{m,2}\,, \\
  \left({\cal V}_{m_1,2}\otimes{\cal R}_{m_2,1}\right)_{\mathrm f} &=& 
    \bigoplus_m \left({\cal R}_{m,2}\oplus
    \underline{{\cal V}_{m-1,3}}
    \oplus{\cal V}_{m+1,3}\right)\,, \\
  \left({\cal V}_{m_1,2}\otimes{\cal R}_{m_2,2}\right)_{\mathrm f} &=& 
    \bigoplus_m \left({\cal R}_{m,1}\oplus2{\cal V}_{m,3}\right)\,, \\
  \left({\cal V}_{m_1,3}\otimes{\cal R}_{m_2,1}\right)_{\mathrm f} &=& 
    \bigoplus_m \left(2{\cal V}_{m,3}\oplus
    \underline{{\cal R}_{m-1,2}}
    \oplus{\cal R}_{m+1,2}\right)\,, \\
  \left({\cal V}_{m_1,3}\otimes{\cal R}_{m_2,2}\right)_{\mathrm f} &=& 
    \bigoplus_m \left(2{\cal R}_{m,2}\oplus
    \underline{{\cal V}_{m-1,3}}
    \oplus{\cal V}_{m+1,3}\right)\,, \\[\medskipamount]
  \left({\cal R}_{m_1,1}\otimes{\cal R}_{m_2,1}\right)_{\mathrm f} &=&
    \bigoplus_m \Bigl( 2{\cal R}_{m,1}\oplus
    \underline{{\cal R}_{m-1,2}}
    \oplus{\cal R}_{m+1,2}\oplus{\cal V}_{m,3}\oplus 
    \nonumber\\*&&\qquad
    \underline{\underline{{\cal V}_{m-2,3}}}\oplus
    \underline{{\cal V}_{m,3}}\oplus{\cal V}_{m+2,3}\Bigr)\,, \\
  \left({\cal R}_{m_1,1}\otimes{\cal R}_{m_2,2}\right)_{\mathrm f} &=&
    \bigoplus_m \left( 2{\cal R}_{m,2}\oplus
    \underline{{\cal R}_{m-1,1}}
    \oplus{\cal R}_{m+1,1}\oplus2\underline{{\cal V}_{m-1,3}}
    \oplus2{\cal V}_{m+1,3}\right)\,, \\
  \left({\cal R}_{m_1,2}\otimes{\cal R}_{m_2,2}\right)_{\mathrm f} &=&
    \bigoplus_m \left( 2{\cal R}_{m,1}\oplus
    \underline{{\cal R}_{m-1,2}}
    \oplus{\cal R}_{m+1,2}\oplus4{\cal V}_{m,3}\right)\,,
\end{eqnarray*}
where the underlined terms on the right hand side are absent for $m=1$
(which can only happen for $m_1=m_2$), and the doubly underlined term
is absent for $m=1,2$ (which only happens for $|m_1 - m_2|\leq1$). 

This shows that the set of representations ${\cal R}_{m,n}$ and 
${\cal V}_{m,n}$ is closed under fusion.

\subsection{$t\geq4$}

In Section \ref{sec:reps} we conjectured (\ref{eq:decomposition}) 
the decomposition of the fusion product of arbitrary 
{\em irreducible\/} representations for all integer $t$.  
The structure of fusion products involving indecomposable
representations is more complicated, and {\em a priori\/} 
one cannot exclude the possibility of generating new types of
representations. However, we have calculated the products
\begin{eqnarray}
  ({\cal V}_{1,2}\otimes{\cal R}_{1,1})_{\mathrm f} &=& 
  {\cal R}_{1,2} \oplus {\cal V}_{2,t}, \nonumber\\*
  ({\cal V}_{1,2}\otimes{\cal R}_{1,n})_{\mathrm f} &=& 
  {\cal R}_{1,n-1} \oplus {\cal R}_{1,n+1}, \qquad\mbox{for
    $1<n<t-1$}, \\
  ({\cal V}_{1,2}\otimes{\cal R}_{1,t-1})_{\mathrm f} &=& 
  {\cal R}_{1,t-2} \oplus 2{\cal V}_{1,t} \nonumber
\end{eqnarray}
for $t=4,\ldots,10$ (up to fusion level $L=2$), and we conjecture that
these identities hold in general. This is then already enough to
derive the fusion rules for all representations, using the
associativity and symmetry of the fusion product. In particular, we
see that the set of representations ${\cal R}_{m,n}$ and
${\cal V}_{m,n}$ is closed under fusion.  

We should note that for all $t$, the subset of representations
${\cal V}_{m,1}$ is closed under fusion, and that the conformal weights
of the fields corresponding to ${\cal V}_{\mathrm{odd},1}$ are all
integral. In particular, the chiral algebra can thus be extended to
include these fields. For odd $t$, there is also another closed subset
of representations given by ${\cal R}_{\mathrm{odd},\mathrm{odd}},   
{\cal R}_{\mathrm{even},\mathrm{even}}$ and 
${\cal V}_{\mathrm{odd},t}$.

\section{Conclusions}
\label{sec:disc}

In this paper we have analysed the fusion products of certain
representations of the $(1,q)$ models. In particular, we have focused
on the fusion products which are not completely reducible, and we have
analysed the structure of the indecomposable representations in
detail. We have determined the parameter which characterises the
indecomposable representations for a few cases explicitly. We have
also studied the fusion rules of these indecomposable representations,
and we have seen that a suitable extended set of representations
closes under fusion. 

This extended set of representations is the smallest set of
representations (containing all allowed irreducible representations)
which closes under fusion. It is therefore the natural analogue of a
``minimal model'' for the $(1,q)$ case, and it seems plausible that it
should give rise to a consistent conformal field theory. In order to
check whether this is indeed the case, it would be necessary to
determine the (chiral) correlation functions, and to analyse whether
there exist suitable chiral-antichiral combinations which define a
crossing-symmetric and local theory.

In the same spirit, it would be interesting to analyse these
representations from the point of view of an enlarged chiral symmetry
algebra. In particular, one could consider the extension of the
Virasoro algebra by three fields of conformal weight $(2q -1)$
(corresponding to the field $(3,1)$), the so-called ``triplet
algebra'' \cite{Kausch91}. It might then happen that the infinitely
many representations in the extended set (which close under fusion
with respect to the Virasoro algebra) form finitely many
representations of the triplet algebra (which close under fusion with
respect to the triplet algebra). The corresponding theory would then 
be a ``rational logarithmic'' conformal field theory. This is 
currently under investigation.
\smallskip

We have also introduced a new algorithm which allows a finite analysis
of fusion products up to any given level. The algorithm can in
principle be applied to any chiral theory, but we have only been able
to show that it terminates for the case of the Virasoro algebra. It
would be interesting to show that this restriction is not necessary.

\paragraph{Acknowledgements}
We would like to thank Wolfgang Eholzer, Michael Flohr, Peter Goddard,
Werner Nahm, Falk Rohsiepe and G\'erard Watts for useful discussions. 
\smallskip

M.R.G. is supported by a Research Fellowship of Jesus College,
Cambridge, and H.G.K. by a Research Fellowship of Sidney Sussex
College, Cambridge. This work has also been supported in part by
PPARC. 

The computer calculations were performed on computers purchased on
EPSRC grant GR/J73322, using a MAPLE package written by H.G.K.

\section*{Appendix: An example}

As an example, we present here the matrices obtained for the fusion
product
\begin{displaymath}
  \left( {\cal V}_{2,1}\otimes{\cal V}_{1,2}\right)_{\mathrm f}
     = {\cal R}_{2,1}
\end{displaymath}
for $t=2$ up to level four. The canonical bases for 
${\cal R}_{2,1}^n$ up to level $L=4$ are composed of four blocks
consisting of lexicographically ordered monomials on the states $\xi,
\phi, \psi$ and $\rho$,  

\begin{displaymath}
\renewcommand{\arraystretch}{1.2}
  \begin{array}{rccccc}
    \hbox{level:}&0&1&2&3&4 \\\hline
    {}[\xi]\colon &
    \xi\,, && L_{-2}\xi\,, & L_{-3}\xi\,, & L_{-4}\xi\,, L_{-2}^2\xi\,, 
    \\ 
    {}[\phi]\colon &&
    \phi\,, & L_{-1}\phi\,, & L_{-2}\phi\,, & L_{-3}\phi\,, 
    L_{-2}L_{-1}\phi\,, 
    \\ 
    {}[\psi]\colon &
    \psi\,, & L_{-1}\psi\,, & L_{-2}\psi\,, & L_{-3}\psi\,, 
    L_{-2}L_{-1}\psi\,, & 
    L_{-4}\psi\,, L_{-3}L_{-1}\psi\,, L_{-2}^2\psi\,, 
    \\ 
    {}[\rho]\colon &&&
    \rho\,, & L_{-1}\rho\,, & L_{-2}\rho\,, L_{-1}^2\rho\,,
  \end{array}
\end{displaymath}
where, for notational simplicity, we omitted the index $(m,n)=(2,1)$
on the vectors. 
In our convention we have
\begin{eqnarray*}
  \phi &=& L_{-1} \xi, \\
  \rho &=& (L_{-1}^2 - 2L_{-2}) \psi - \frac12 L_{-3} \xi.
\end{eqnarray*}
In the following, we give the matrices for 
\begin{displaymath}
  L_n\colon {\cal R}_{2,1}^4 \to {\cal R}_{2,1}^{4-n}\,, \qquad
  n=0\,,\ldots4\,, 
\end{displaymath}
and indicate the four blocks of the basis. The matrix for $L_0$ is
given as
\begin{displaymath}
  L_0 = 
  \begin{array}{r@{}r@{}ccccc|ccccc|cccccccc|cccc@{}l}
    && \multicolumn{5}{c}{[\xi]} & \multicolumn{5}{c}{[\phi]} & 
    \multicolumn{8}{c}{[\psi]} & \multicolumn{4}{c}{[\rho]} \\
    &&
    0& .& .& .& .& .& .& .& .& .&
    .& .& .& .& .& .& .& .& .& .& .& .\\
    &&
    .& 2& .& .& .& .& .& .& .& .&
    .& .& .& .& .& .& .& .& .& .& .& .\\
    &&
    .& .& 3& .& .& .& .& .& .& .&
    .& .& .& .& .& .& .& .& .& .& .& .\\
    &&
    .& .& .& 4& .& .& .& .& .& .&
    .& .& .& .& .& .& .& .& .& .& .& .\\
    \multirow5{[\xi]}&&
    .& .& .& .& 4& .& .& .& .& .&
    .& .& .& .& .& .& .& .& .& .& .& .\\
    \cline{3-24}
    &&
    .& .& .& .& .& 1& .& .& .& .&
    1& .& .& .& .& .& .& .& .& .& .& .\\
    &&
    .& .& .& .& .& .& 2& .& .& .&
    .& 1& .& .& .& .& .& .& .& .& .& .\\
    &&
    .& .& .& .& .& .& .& 3& .& .&
    .& .& 1& .& .& .& .& .& .& .& .& .\\
    &&
    .& .& .& .& .& .& .& .& 4& .&
    .& .& .& 1& .& .& .& .& .& .& .& .\\
    \multirow5{[\phi]}&&
    .& .& .& .& .& .& .& .& .& 4& .& .& .& .& 1& .& .& .& .& .& .& .\\
    \cline{3-24}
    &&
    .& .& .& .& .& .& .& .& .& .&
    1& .& .& .& .& .& .& .& .& .& .& .\\
    &&
    .& .& .& .& .& .& .& .& .& .&
    .& 2& .& .& .& .& .& .& .& .& .& .\\
    &&
    .& .& .& .& .& .& .& .& .& .&
    .& .& 3& .& .& .& .& .& .& .& .& .\\
    &&
    .& .& .& .& .& .& .& .& .& .&
    .& .& .& 4& .& .& .& .& .& .& .& .\\
    &&
    .& .& .& .& .& .& .& .& .& .&
    .& .& .& .& 4& .& .& .& .& .& .& .\\
    &&
    .& .& .& .& .& .& .& .& .& .&
    .& .& .& .& .& 5& .& .& .& .& .& .\\
    &&
    .& .& .& .& .& .& .& .& .& .&
    .& .& .& .& .& .& 5& .& .& .& .& .\\
    \multirow8{[\psi]}&&
    .& .& .& .& .& .& .& .& .& .&
    .& .& .& .& .& .& .& 5& .& .& .& .\\
    \cline{3-24}
    &&
    .& .& .& .& .& .& .& .& .& .&
    .& .& .& .& .& .& .& .& 3& .& .& .\\
    &&
    .& .& .& .& .& .& .& .& .& .&
    .& .& .& .& .& .& .& .& .& 4& .& .\\
    &&
    .& .& .& .& .& .& .& .& .& .&
    .& .& .& .& .& .& .& .& .& .& 5& .\\
    \multirow4{[\rho]}&\multiparen{22}(.&
    .& .& .& .& .& .& .& .& .& .&
    .& .& .& .& .& .& .& .& .& .& .& 5
    & \multiparen{22}.) \\
  \end{array}
\end{displaymath}
This matrix explicitly exhibits the Jordan cell structure between the
blocks $[\phi]$ and $[\psi]$. 

According to our discussion of Section \ref{sec:reps}, the matrices
for the positive Virasoro modes have the general structure 
\begin{displaymath}
  \begin{array}{r@{}r@{}cccc@{}l}
    &&[\xi]&[\phi]&[\psi]&[\rho]\\
    {}[\xi] &&*&0&*&0 \\
    {}[\phi]&&*&*&*&* \\
    {}[\psi]&&0&0&*&0 \\
    {}[\rho]&\multiparen4(.&0&0&*&* & \multiparen4.) 
  \end{array}
\end{displaymath}
reflecting, for example, that states in the block $[\xi]$ cannot be
mapped to states in the blocks $[\psi]$ and $[\rho]$, and similarly in
the other cases. In more detail, we find 
\begin{displaymath}
  L_1 = \!
  \begin{array}{r@{}r@{}ccccc|ccccc|cccccccc|cccc@{}l@{}}
    && \multicolumn{5}{c}{[\xi]} & \multicolumn{5}{c}{[\phi]} & 
    \multicolumn{8}{c}{[\psi]} & \multicolumn{4}{c}{[\rho]} \\
    &&
    .& .& .& .& .& .& .& .& .& .&\fbox{$-1$}& .& .& .& .& .& .& .& .& .& .& .\\
    &&
    .& .& 4& .& .& .& .& .& .& .& .& .& -1& .& .& .& .& .& .& .& .& .\\
    \multirow3{[\xi]}&&
    .& .& .& 5& 3& .& .& .& .& .& .& .& .& -1& 3/2& .& .& .& .& .& .& .\\
    \cline{3-24}
    &&
    .& 3& .& .& .& .& 2& .& .& .& .& 1& .& .& .& .& .& .& .& .& .& .\\
    &&
    .& .& .& .& .& .& .& 3& .& .& .& .& .& .& .& .& .& .& 3& .& .& .\\
    \multirow3{[\phi]}&&
    .& .& .& .& 6& .& .& .& 4& 8& .& .& .& .& 1& .& .& .& .& 6& .& .\\
    \cline{3-24}
    &&
    .& .& .& .& .& .& .& .& .& .& .& 2& .& .& .& .& .& .& .& .& .& .\\
    &&
    .& .& .& .& .& .& .& .& .& .& .& .& 3& .& .& .& .& .& .& .& .& .\\
    &&
    .& .& .& .& .& .& .& .& .& .& .& .& .& 4& 8& .& .& .& .& .& .& .\\
    &&
    .& .& .& .& .& .& .& .& .& .& .& .& .& .& .& 5& 2& 3& .& .& .& .\\
    \multirow5{[\psi]}&&
    .& .& .& .& .& .& .& .& .& .& .& .& .& .& .& .& 4& 6& .& .& .& .\\
    \cline{3-24}
    &&
    .& .& .& .& .& .& .& .& .& .& .& .& .& .& 3& .& .& .& .& 6& .& .\\
    \multirow2{[\rho]}&\multiparen{13}(.&
    .& .& .& .& .& .& .& .& .& .& .& .& .& .& .& .& .& .& .& .& 3& 14
    & \multiparen{13}.) \\
  \end{array} ,
\end{displaymath}
where the boxed entry is the characteristic parameter,
$\beta_{2,1}=-1$, of ${\cal R}_{2,1}$. The other matrices are: 
\begin{displaymath}
  L_2 = \!\!
  \begin{array}{@{}r@{}r@{}ccccc|ccccc|cccccccc|cccc@{}l@{}}
    && \multicolumn{5}{c}{[\xi]} & \multicolumn{5}{c}{[\phi]} & 
    \multicolumn{8}{c}{[\psi]} & \multicolumn{4}{c}{[\rho]} \\
    &&
    .& -1& .& .& .& .& .& .& .& .& .& -3& .& .& .& .& .& .& .& .& .& .\\
    \multirow2{[\xi]}&&
    .& .& .& 6& 6& .& .& .& .& .& .& .& .& .& -3& .& .& .& .& .& .& .\\
    \cline{3-24}
    &&
    .& .& 5& .& .& .& .& 3& .& .& .& .& 4& .& .& .& .& .&
    \hbox to1em{\hss$-\frac{21}{2}$\hss}& .& .& .\\
    \multirow2{[\phi]}&&
    .& .& .& .& .& .& .& .& 5& 7& .& .& .& .& 4& .& .& .&
    .& \hbox to1em{\hss$-\frac32$\hss}& .& .\\
    \cline{3-24}
    &&
    .& .& .& .& .& .& .& .& .& .& .& .& 3& .& .& .& .& .& .& .& .& .\\
    &&
    .& .& .& .& .& .& .& .& .& .& .& .& .& 5& 7& .& .& .& .& .& .& .\\
    \multirow3{[\psi]}&&
    .& .& .& .& .& .& .& .& .& .& .& .& .& .& .& 6& 10& 14& .& .& .& .\\
    \cline{3-24}
    {}[\rho]&\multiparen8(.&
    .& .& .& .& .& .& .& .& .& .& .& .& .& .& .& .& 5& .& .& .& 11& 18
    &\multiparen8.)\\
    &&\multicolumn{10}{c}{}&\multicolumn{1}{c}{\Uparrow}&
    \multicolumn{11}{c}{}
  \end{array}
\end{displaymath}
\begin{displaymath}
  L_3 = \!
  \begin{array}{@{}r@{}r@{}ccccc|ccccc|cccccccc|cccc@{}l@{}}
    && \multicolumn{5}{c}{[\xi]} & \multicolumn{5}{c}{[\phi]} & 
    \multicolumn{8}{c}{[\psi]} & \multicolumn{4}{c}{[\rho]} \\
    {}[\xi]&&
    .& .& -4& .& .& .& .& .& .& .& .& .& -5& .& .& .& .& .& .& .& .& .\\ 
    \cline{3-24}
    {}[\phi]&&
    .& .& .& 7& 15& .& .& .& 2& 10& .& .& .& 6& 5& .& .& .& .& -42& .& .\\ 
    \cline{3-24}
    &&
    .& .& .& .& .& .& .& .& .& .& .& .& .& 2& 10& .& .& .& .& .& .& .\\
    \multirow2{[\psi]}&\multiparen{4}(.&
    .& .& .& .& .& .& .& .& .& .& .& .& .& .& .& 7& 8& 15& .& .& .& .
    &\multiparen4.)\\
    &&\multicolumn{10}{c}{}&\multicolumn{1}{c}{\Uparrow}&
    \multicolumn{11}{c}{}
  \end{array}
\end{displaymath}
\begin{displaymath}
  L_4 = \!
  \begin{array}{@{}r@{}r@{}ccccc|ccccc|cccccccc|cccc@{}l@{}}
    && \multicolumn{5}{c}{[\xi]} & \multicolumn{5}{c}{[\phi]} & 
    \multicolumn{8}{c}{[\psi]} & \multicolumn{4}{c}{[\rho]} \\
    {}[\xi]&&
    .& .& .& -10& -6& .& .& .& .& .& .& .& .& -7& -18& .& .& .& .& .& .& .\\ 
    \cline{3-24}
    {}[\psi]&\multiparen{2}(.&
    .& .& .& .& .& .& .& .& .& .& .& .& .& .& .& -2& 14& 18& .& .& .& .
    &\multiparen2.)\\
    &&\multicolumn{10}{c}{}&\multicolumn{1}{c}{\Uparrow}&
    \multicolumn{11}{c}{}
  \end{array}
\end{displaymath}
Here we have indicated by an arrow the column corresponding to 
$L_n \psi = 0$ for $n>1$.

\end{document}